\documentclass[usegraphicx,usenatbib,useapjfonts,apj]{emulateapj}
\usepackage{amssymb}
\usepackage{aas_macros}
\usepackage{graphicx} % Include figure files

\usepackage{psfrag,color}
\usepackage{psfig}
\usepackage{times}

\usepackage{longtable}

\def\cm3{\mbox{cm$^{-3}$}}
\def\erg{\mbox{ergs}}
\def\Ke{\mbox{K}}

\def\kpch{\mbox{$h^{-1}$kpc}}

\def\lcdm{{$\Lambda$CDM}}
\def\mpch{\mbox{$h^{-1}$Mpc}}

\def\msun{\mbox{M$_\odot$}}

\def\msunh{\mbox{$h^{-1}$M$_\odot$}}    

\def\mh{\mbox{$M_{h}$}}
\def\ms{\mbox{$M_{s}$}}
\def\mstar{\mbox{$M^{*}$}}
\def\mg{\mbox{$M_{g}$}}
\def\mb{\mbox{$M_{b}$}}
\def\fg{\mbox{$f_{g}$}}
\def\fs{\mbox{$F_{s}$}}
\def\fb{\mbox{$F_{b}$}}
\def\fbU{\mbox{$F_{b,U}$}}

\def\ome{\mbox{$\Omega_0$}}
\def\omel{\mbox{$\Omega_\Lambda$}}
\def\omeb{\mbox{$\Omega_b$}}

\def\rv{\mbox{$R_{\rm vir}$}}
\def\re{\mbox{$R_{e}$}}

\def\nbody{\mbox{$N-$body}}

\def\fsf{\mbox{$\epsilon_{\rm SF}$}}
\def\vmax{\mbox{$V_{\rm max}$}}

\def\ltsima{$\; \buildrel < \over \sim \;$}    % Use in text mode
\def\lesssim{\lower.5ex\hbox{\ltsima}}           % Use in math mode
\def\gtsima{$\; \buildrel > \over \sim \;$}    % Use in text mode
\def\gtrsim{\lower.5ex\hbox{\gtsima}}           % Use in math mode

\newcommand{\Dvir}{\Delta_{\rm vir}}
\newcommand{\kms}{{\rm ~km~s}^{-1}}
\newcommand{\lmax}{\ell_{\rm max}}

\newcommand{\mpart}{m_{{\rm dm}}}

\newcommand{\nsf}{n_{\rm SF}}

\newcommand{\pcc}{{\rm cm}^{-3}}

\newcommand{\Tsf}{T_{\rm SF}}
\newcommand{\toff}{t_{\rm off}}

\def\spose#1{\hbox to 0pt{#1\hss}}
\newcommand\lsim{\mathrel{\spose{\lower 3pt\hbox{$\mathchar"218$}}
     \raise 2.0pt\hbox{$\mathchar"13C$}}}
\newcommand\gsim{\mathrel{\spose{\lower 3pt\hbox{$\mathchar"218$}}
     \raise 2.0pt\hbox{$\mathchar"13E$}}}
     
\def\apj{\mbox{ApJ}}
\def\apjs{\mbox{ApJSS}}
\def\mnras{\mbox{MNRAS}}
\def\apjl{\mbox{ApJL}}

\def\aj{\mbox{AJ}}
\def\aap{\mbox{A\&A}}

\slugcomment{Submitted}

\shorttitle{Cosmological simulations of low-mass central galaxies}
\shortauthors{Avila-Reese et al.}

\begin{document}

\title{The specific star formation rate and stellar mass fraction of low-mass central galaxies in cosmological simulations}
\author{V. Avila-Reese\altaffilmark{1}, P. Col\'{\i}n\altaffilmark{2}, A. Gonz\'alez-Samaniego\altaffilmark{1},
O. Valenzuela\altaffilmark{1}, C. Firmani\altaffilmark{1,3}, 
H. Vel\'azquez\altaffilmark{4}, and D. Ceverino\altaffilmark{5}
}

\altaffiltext{1}{Instituto de Astronom\'{\i}a, Universidad Nacional Aut\'onoma de M\'exico, 
A.P. 70-264, 04510, M\'exico, D.F., M\'exico}

\altaffiltext{2}{Centro de Radioastronom\'{\i}a y Astrof\'{\i}sica, Universidad Nacional 
Aut\'onoma de M\'exico, A.P. 72-3 (Xangari), Morelia, Michoac\'an 58089, M\'exico }

\altaffiltext{3}{INAF-Osservatorio Astronomico di Brera, via E.Bianchi 46, I-23807 Merate, Italy}

\altaffiltext{4}{Instituto de Astronom\'{\i}a, Universidad Nacional Aut\'onoma de M\'exico, A. P. 877, 
Ensenada BC 22800, M\'exico}

\altaffiltext{5}{Racah Institute of Physics, The Hebrew University, Jerusalem 91904, Israel}

\begin{abstract} 

By means of cosmological \nbody\ + Hydrodynamics simulations of galaxies
in the context of the $\Lambda$ Cold Dark Matter (\lcdm) scenario we explore the 
specific star formation rates (SSFR=SFR/\ms, \ms\ is the stellar mass) 
and stellar mass fractions ($\fs\equiv \ms/\mh$, \mh\ is the halo mass)
for sub-\mstar\ field galaxies at different redshifts ($0\lesssim z\lesssim 1.5$). 
Distinct low-mass halos ($2.5\lesssim \mh/10^{10} \msun \lesssim 50$ at $z=0$)
were selected for the high-resolution re-simulations. The Hydrodynamics 
Adaptive Refinement Tree (ART) code was used and some variations of the
sub-grid parameters were explored. Most simulated galaxies, specially those
with the highest resolutions, have significant disk components and their structural 
and dynamical properties are in reasonable agreement with observations of 
sub-\mstar\ \textit{field} galaxies. However, the SSFRs are 5-10 times smaller than 
the averages of several (compiled and homogenized here) observational determinations 
for field blue/star-forming galaxies at $z<0.3$ (at low masses, most observed field 
galaxies are actually blue/star-forming). This inconsistency seems to remain even at 
$z\sim 1-1.5$, although it is less drastic. The \fs\ of simulated galaxies increases with \mh\ 
as semi-empirical inferences show. However, the values of \fs\ at $z\approx 0$ 
are $\sim 5-10$ times larger in the simulations than in the inferences; these 
differences increases probably to larger 
factors at $z\sim 1-1.5$.  The inconsistencies reported here imply 
that simulated low-mass galaxies ($0.2\lesssim \ms/10^{9} \msun \lesssim 30$ at $z=0$) 
assembled their stellar masses much earlier than observations suggest. 
Our results confirm the predictions found by means of \lcdm-based models of disk galaxy 
formation and evolution for isolated low-mass galaxies \citep{FA10}, and highlight that our 
understanding and implementation of astrophysics into simulations and models are still 
lacking vital ingredients. 
\end{abstract}


%=====================
\section{Introduction}
%=====================

Based on the recent observational achievements, an empirical 
picture of stellar mass (\ms) assembly history of galaxies 
 as a function of mass is emerging. Probably, the most remarkable 
 aspect in this picture is the so-called 'downsizing' phenomenon.
The term 'downsizing' was coined by \cite{Cowie96} to describe the rapid
decline with cosmic time of the maximum rest-frame $K-$band luminosity of 
galaxies undergoing active star formation (SF).  Since then, as new observational
inferences appeared, this term has been used to describe a number of trends 
of the galaxy population as a function of mass, most of them related actually to 
different astrophysical phenomena and galaxy evolutionary stages as discussed 
in \citet[][see also Neistein et al. 2006; Santini et al. 2009]{Fontanot09} .

 From the most general point of view, the many downsizing manifestations can be separated
 into those that refer to the evolution (i) of massive galaxies, which today are on average 
 red and passive (quenched SF), and (ii) of less massive galaxies, which today are 
 on average blue and star forming. This separation seems to have an astrophysical root;
that is, it is not based on a merely methodological division. For example, 
\citet[][hereafter FA10]{FA10} have inferred the \textit{average} \ms\ growth tracks of 
galaxies as a function of mass (the 'galaxian hybrid evolutionary tracks', GHETs) 
 by means of a semi-empirical approach, and showed that at each
 epoch there is a characteristic mass that separates galaxies into two populations.
 Galaxies more massive than \ms($z=0$)$\approx 3\times 10^{10}$ \msun\ 
 are on average passive (their SSFRs have dramatically diminished,
 where SSFR=SFR/\ms\ is the specific SF rate), besides the more massive
 is the galaxy, the earlier it transited from the active (blue, star-forming) to the passive
 (red, quenched) population ('population downsizing'). Galaxies less massive than 
 \ms($z=0$)$\approx 3\times 10^{10}$ \msun\ are on average 
 yet blue and active, and the less massive the galaxy,  the higher its SSFR and/or the 
 faster its late \ms\ growth ('downsizing in SSFR').
 We recall that these are just average trends. In fact, there are other factors besides mass
 that intervene in the stellar mass build-up of galaxies, e.g., the large-scale environment in 
 which galaxies evolve \citep[e.g.,][and more references therein]{Peng10} and whether 
 they are central or satellites objects \citep[for a recent review see][]{Weinmann11}. 
 
 A large amount of {\it direct look-back time} observations support the mentioned 
 two downsizing phenomena:
 \begin{itemize}
 \item {\it For high-mass  galaxies,} those with $\ms\gtrsim \mstar$, where
\mstar\ is the characteristic mass of the stellar mass function ($\approx10^{10.8}$ \msun\
at $z\sim 0$), observations show the existence of a decreasing with cosmic time characteristic mass  
 at which the SFR is dramatically quenched or at which the stellar mass functions of 
 early- and late-type galaxies cross, evidencing this a mass-dependent migration from
 blue to red population (e.g., Bell et al. 2003,2007; Bundy et al. 2006; Hopkins et al. 2007; 
 Drory \& Alvarez 2008; Vergani et al. 2008; Pozzetti et al. 2010).

\item {\it For low-mass galaxies} ($\ms<< \mstar$, they are mostly late-type, star-forming 
systems), at least up to $z\sim 1-2$, and by using different SFR tracers and methods to 
estimate \ms, it was found that their SSFRs are surprisingly high even at $z\sim 0$ and, 
on average, the lower the mass, the higher the SSFR \citep[e.g.,][]{B04,Bauer05,Feulner+05,Zheng07,Noeske07,Bell07,Elbaz07,Salim07,Chen09,Damen09a,Damen09b,
Santini09,Oliver10,Rodighiero10,Karim10}.  
\end{itemize}

\subsection{Confronting  the empirical picture to theoretical predictions}

The current paradigm of galaxy formation and evolution is based on the hierarchical clustering
scenario \citep{WR78,WF91} within the context of the $\Lambda$ Cold Dark Matter (\lcdm) cosmology. 
According to this scenario, galaxies form and evolve in the centres of hierarchically growing \lcdm\
halos.  Though at first glance contradictory, it was shown that the 'population (or archaelogical) downsizing' 
related to massive galaxies has, at least partially,  its natural roots in the hierarchical clustering 
process of the dark-matter halos and their progenitors distribution 
\citep[][see also Guo \& White 2008; Li, Mo \& Gao 2008; Kere{\v s} et al. 2009]{Neisten06}.  
Besides, red massive galaxies are expected to have been formed in early collapsed 
massive (associated to high peak and clustered) halos that afterwards become part of groups 
and clusters of galaxies, leaving truncated therefore the mass growth of the galaxies associated 
to these halos.  On the other hand, massive galaxies typically hosted in the past active galactic 
nuclei (AGN). The strong feedback of the AGN may help to stop gas accretion,
truncating further the galaxy stellar growth and giving rise to shorter formation time-scales for more 
massive galaxies \citep{Bower06,Croton06,DeLucia06}.

All the factors mentioned above work in the direction of reproducing the downsizing 
manifestations of massive galaxies within the hierarchical \lcdm\ scenario \citep[]{Fontanot09}
--though several questions remain yet open.

What about the downsizing related to low-mass galaxies?
As discussed in \citet{FAR10} and FA10, disk galaxy evolutionary 
models in the context of the hierarchical \lcdm\ scenario seem to face difficulties in reproducing 
both the high empirically determined values of SSFR and the SSFR downsizing trend of low-mass
galaxies ($\ms\lesssim 3\times 10^{10} \msun\ <<\mstar$),
which are mainly blue/star-forming systems of disk-like morphology. 

The gas infall rate onto the (model) disks is primarily driven by the cosmological halo
mass aggregation history \citep[]{FA00,vandenBosch00,Stringer07,Dutton09,FAR10,
Dutton10a}, which is hierarchical, i.e. the less massive the halo, the earlier its fast 
mass aggregation rate phase. The gas infall rate is the main driver of disk SFR 
(other factors related to the process of gas conversion into stars, local feedback, 
interstellar medium turbulence, etc. introduce minor systematic deviations). The 
large-scale feedback over the gas exerted mainly by SNe explosions, may strongly 
affect the SFR history of low-mass disks mainly because it alters the primary gas infall rate.  
The SN feedback has been commonly invoked in semi-analytical models 
(SAM) to reproduce the faint-end flattening of the galaxy luminosity (or \ms) function, 
assuming that this feedback produces gas reheating and galactic outflows 
\citep[see e.g.,][and more references therein ]{Benson03}.  \citet{FAR10} experimented 
with different models of SN-driven galactic outflows (e.g., based on energy and 
momentum conservation) besides of the disk gas turbulence input due to SN 
feedback. They have found that the present-day stellar and baryonic mass fractions
($\fs\equiv\ms/\mh$ and $\fb\equiv\mb/\mh$, where \mh\ is the total -virial- halo mass 
and \mb = \ms + \mg, \mg\ is the galaxy cold gas mass) of low-mass galaxies can 
be roughly reproduced when assuming extreme (probably unrealistic) galactic 
outflow efficiencies \citep[see also][]{Dutton09,Dutton10a}. However, in any model
with galactic outflows it was possible to alter the SFR histories of their disk 
galaxies in order to reproduce the observed (at least up to $z\sim 1$) downsizing 
in SSFR and the too high SSFRs measured for small central galaxies at late epochs.  

\citet{FAR10} experimented also with the possibility of late re-accretion of the ejected gas
\citep[in previous works it was assumed that the feedback-driven outflows eject gas 
from the small halos forever, but see e.g.,][]{Springel01,deLucia04,Bertone07,Oppenheimer08,
Oppenheimer10}. For reasonable schemes of gas re-accretion as a function of halo 
mass, \citet{FAR10}  found that the SSFR of galaxies increases but it does it in 
the opposite direction of the downsizing trend:  the increase is larger for the more 
massive galaxies.

A related issue of \lcdm-based low-mass (disk) galaxy models appears in the 
evolution of the \ms--\mh\ (or \fs--\mh) relation. The semi-empirical inferences of this 
relation \citep[see e.g.,][]{CW09,Moster10,WangJing10,BCW10} show that for a fixed 
value of \mh, \fs\ is not only very small at $z=0$, but it becomes smaller at higher $z$. 
The \lcdm-based disk galaxy evolutionary models predict the opposite, i.e. higher values 
of \fs\ at higher $z$ for a fixed value of \mh\ (FA10). 
%By connecting the empirical \ms--\mh\ 
%relations with individual average halo mass aggregation histories (MAHs), FA10 
%\citep[see also][]{CW09} calculated the corresponding \textit{individual} (average) \ms\ 
%(or \fs) evolutionary tracks. These tracks imply a much faster growth of \ms\ (\fs) than \mh\ 
%since $z\sim 1-2$, while for modeled galaxies \ms\ (\fs) grows only slightly faster than \mh.  

The inconsistence between models and observations related to the stellar 
mass build-up of sub-\mstar\ disk galaxies found in \citet{FAR10} and FA10 
is connected with some issues reported in recent SAMs. In these works, 
based also on the hierarchical \lcdm\ scenario, it was showed that the stellar 
population of small (both central and satellite) galaxies ($\ms\approx 10^{9.0}-10^{10.5}$ \msun) 
is assembled too early, becoming these galaxies older, redder, and with 
lower SSFRs at later epochs than the observed galaxies in the same mass 
range \citep{Somerville08, Fontanot09, Santini09,Pasquali10,Liu10}. 
By means of a disk galaxy evolutionary model similar to FA10, \citet{Dutton10a} 
found also that the SSFR of low-mass galaxies is below the average
of that given by observations, specially at $z\sim 1-2$, though these authors 
conclude that their models roughly reproduce the main features of the observed
SFR sequence.   

\subsection{Cosmological simulations of galaxy evolution}

Due to the high non-linearity implied in the problem of halo/galaxy formation and evolution, 
cosmological N-body/hydrodynamical simulations offer the fairest way to attain
a realistic modeling of galaxies. 
However, this method is hampered by the large --currently unreachable-- 
dynamic range required to model galaxy formation and evolution in 
the cosmological context,
as well as by the complexity of the processes involved, mainly gas thermo-hydrodynamics 
and SF and its feedback on the surrounding medium. The latter processes occur 
at scales ($\lesssim 1-50$ pc) commonly below the accessible resolution in simulations.
Therefore, "sub-grid" schemes based on physical models for these processes
should be introduced in the codes. 

In the last years, cosmological simulations have matured enough as to produce
individual galaxies that at $z\sim 0$ look quite realistic \citep[e.g.,][]{Governato07,
Governato10,Naab07,Mayer08,Zavala08,CK2009,Gibson09,Scannapieco08,
Piontek09,Colin10,Sawala10,Agertz11}, 
though this success in most cases is conditioned by the assumed sub-grid 
schemes and parameters, and the resolution that limits the simulations to
small boxes, where only one or a few galaxies are followed.

{\it What do current cosmological N-body/hydrodynamical simulations 
of low-mass galaxies predit regarding their SSFR and \fs\ evolution?} By using 
the Adaptive Refinement Tree code \citep[ART][]{KKK97} with hydrodynamics included 
\citep{Kravtsov03}, \citet[][hereafter C10]{Colin10} have explored different SF schemes and sub-grid 
parameters for one simulated low-mass galaxy ($\mh\approx 7\times 10^{10}$ \msun\ 
at $z=0$) that develops a significant disk. 
For some cases, the obtained galaxy did not look like realistic. For other cases, several 
global dynamical, structural, and ISM  properties of observed small galaxies were 
reproduced. However, even in these cases 
the SSFR and \fs\ at different $z$ are respectively much lower and higher
than the observational inferences, an issue that apparently would also 
share other simulations presented in the literature.

In this paper, our aim is {\it to measure at different epochs the SSFRs and 
stellar mass fractions (\fs) of sub-\mstar\ central galaxies covering a large range of
masses} obtained in state-of-the-art N-body/hydrodynamical cosmological simulations, 
and explore whether the mentioned-above issues of \lcdm--based galaxy 
evolutionary models (as well as SAMs) are also present or not in the simulations.
Here we will use the Hydrodynamics ART code with some of the most reliable
sub-grid schemes/parameters explored in C10 and covering almost two orders
of magnitude in \ms. The effects of resolution over the SSFR and \fs\ at different
epochs will be also explored. 
 
The plan of the paper is as follows. The code and the SF/feedback prescriptions used for 
the simulations are described in \S 2. In this section, the cosmological simulation and the 
different runs varying the SF/feedback parameters are also presented. In \S 3.1 some
generalities of the simulated galaxies are discussed, while in \S 3.2 the SSFRs and \fs\ of
all the simulations at $z=0.00, 0.33, 1.00,$ and 1.50 vs mass are presented and 
compared with several observational inferences. In \S\S 4.1  we discuss the numerical
results from previous works, and in \S\S 4.2 the caveats of the observational
determinations used here are discussed. 
Our conclusions are presented in \S 5.  

%=====================
\section{The simulations}
%=====================

\subsection{The Code} \label{sec:the_code}

The numerical simulations used in this work were performed using 
the Hydrodynamics + N-body ART code \citep[]{KKK97,Kravtsov03}.
Among the physical processes included in ART are the cooling of the gas 
and its subsequent conversion into stars, thermal stellar feedback, 
self-consistent advection of metals, a UV heating background source.

The cooling and heating rates incorporate Compton heating/cooling, atomic, and
{\it molecular Hydrogen and metal-lines} cooling, UV heating from a cosmological background 
radiation \citep{HM96}, and are tabulated for a temperature range of 
$10^2 < T < 10^9\ \Ke$ and a grid of densities, metallicities, and redshifts 
using the CLOUDY code \citep[ version 96b4]{Ferland98}. 

Star formation is modeled as taking place in the coldest and densest collapsed regions, 
defined by $T < \Tsf$ and $n_g >\nsf$, where $T$ and $n_g$ are the temperature and 
number density of the gas, respectively, and  $ \Tsf$ and $\nsf$ are a temperature and density 
SF threshold. A stellar particle of mass $m_*$ is placed in a grid cell where these conditions
are simultaneously satisfied, and this mass is removed from the gas mass in the cell. 
The particle subsequently follows N-body dynamics. No other criteria are imposed. 
In most of the simulations presented here, the stellar particle mass, $m_*$, is calculated by
assuming that a given fraction (SF local efficiency factor $\fsf$) of the cell gas mass,
$m_g$, is converted into stars; that is, $m_* = \fsf m_g$, where \fsf\ is
treated as a free parameter. Based on the simulations performed in C10, where
actually \fsf\ was not a fixed parameter but a quantity calculated in each cell
and timestep by an algorithm dependent on other parameters,
we have found that when \fsf\ acquires values around 0.5, the
simulation gives reasonable results. This value
is high enough for the thermal feedback to be efficient in regions
of dense, cold, star-forming gas and not as large as to imply that
most of the cell gas is exhausted. 

In C10, besides the ``deterministic'' SF prescription described above, 
the authors experimented also with a ``stochastic'' or ``random'' SF prescription 
in which stellar particles are created in a cell with a probability function 
proportional to the gas density.  The stochastic prescription allows for the 
possibility (with low probability) of forming stars in regions of low average 
density, in which isolated dense clouds are not resolved. Here, all of our 
simulations, except one, 
use the ``deterministic'' SF prescription assuming 
a density threshold of $\nsf = 6\ \pcc$, which corresponds to a
gas column density $N \sim 10^{21}$ cm$^{-2}$ (roughly the lower limit of 
observed giant molecular clouds) in a cell of $100-150$ pc (see C10).

Since stellar particle masses are much more massive than the mass 
of a star, typically $10^4$ -- $10^5~\msun$, once formed, each stellar particle 
is considered as a single stellar population, within which the individual
stellar masses are distributed according to the Miller \& Scalo IMF.
Stellar particles eject metals and thermal energy through stellar winds and 
type II and Ia supernovae (SNe) explosions. 
Each star more massive than 8 \msun\ is assumed
to dump into the ISM, instantaneously, $2 \times 10^{51}\ \erg$ in the form
of {\it thermal energy};  $10^{51}\ \erg$ comes from the stellar wind, and 
the other $10^{51}\ \erg$ from the SN explosion. Moreover, the star is assumed 
to eject $1.3 \msun$ of metals.
For the assumed \citet{MS79} 
initial mass function, IMF, a stellar particle of $10^5\ \msun$ produces 749 type 
II SNe. For a more detailed discussion of the processes implemented in the code, see 
\citet{Kravtsov03} and \citet{KNV05}.

Stellar particles dump energy in the form of heat to the cells in which they are born. 
Most of this thermal energy, inside the cell, is radiated away unless the cooling is 
turned off temporally.  This mechanism along with a relatively high value of \fsf\  
allow the gas to expand and move away from the star forming region.
%{\bf and deposit not only thermal energy into the surrounding cells but also kinetic energy,
%i.e. a fraction of the stellar wind and SN energy ends as kinetic energy of the ISM .}
Thus, to allow for outflows, in the present paper we adopt the strategy 
of turning off the cooling during a time $\toff$ in the cells where stellar particles form
(see C10). 
As $\toff$ can be linked to the crossing time 
in the cell at the finest grid, we could see this parameter as depending on resolution in the
sense that the higher the resolution, the smaller its value. In most simulations 
presented in this work, we turn off the cooling for $\toff=40$, 20, and 10 Myr depending 
on resolution; however, we have found that varying this time, within this range, for a 
given simulation, does not affect significantly the results.

\subsection{Numerical strategy and the runs}

Most simulations presented here were run in a \lcdm\ cosmology with 
$\ome = 0.3$, $\omel = 0.7$, $\omeb = 0.045$, and $h=0.7$. The CDM power 
spectrum is taken from \citet{kh97} and it is normalized to $\sigma_8 = 0.8$,
where $\sigma_8$ is the rms amplitude of mass fluctuations in 8 \mpch\
spheres. Few simulations were run using a different cosmological set up 
(models C, F, and H of Table 1). For them, $\ome = 0.27$, $\omel = 0.73$, 
$\omeb = 0.047$, and $h= 0.7$, and the power spectrum is that one used to 
run the "Bolshoi simulation" \citep{Klypin10} with $\sigma_8 = 0.82$.
%(on-line code for anisotropies 
%http://lambda.gsfc.nasa.gov/toolbox/tb_camb_form.cfm)

To maximize resolution efficiency, we use the "zoom-in"
technique. First a low-resolution cosmological simulations with 
only dark matter (DM) particles is run, and then regions 
(DM halos) of interest are picked up to be re-simulated with high
resolution and with the physics of the gas included. The low-resolution simulations
were run with $128^3$ particles inside a box of $10 \mpch$ per side,
with the box initially covered by a mesh of $128^3$ cells (zero-th level cells). 
At $z=0$, we search for low-mass halos ($2 \times 10^{10} \lesssim \mh/\msunh 
\lesssim 4 \times 10^{11}$) that are not contained within larger halos ({\it distinct} halos),
but the selection was not based on their MAHs.

Out of ten halos, seven do not have companion halos at $z=0$ with a mass greater 
than 0.2 the mass of the selected halo within a distance  $\lesssim 0.5 \mpch$. On
 the other hand, two of them are 220 \kpch\ away from each other (halos $G$ and $J$, 
see Table 1), and finally, there is one halo ($D$) that is separated by 330 \kpch\
from a halo with a mass of about 0.5 the mass of halo $D$.

A Lagrangian region of 2 or 3\rv\ is identified at $z = 50$ 
and re-sampled with additional small-scale modes \citep{KKBP01}. The 
virial radius, \rv, is defined as the radius that encloses a mean density equal 
to $\Dvir$ times the mean density of the universe, where $\Dvir$
is a quantity that depends on  $\ome$, $\omel$ and $z$. For
example, for our cosmology $\Dvir(z=0) = 338$ and 
$\Dvir(z=1) = 203$.  The number of DM particles in the high-resolution 
region depends on the number of DM species and the mass of the 
halo, but for models with four species (high-resolution) this vary from 
$\sim$ half a million (model Ah) to about 4.7 million (model Ih),
the least massive and the second most massive halos, respectively.
The corresponding dark matter mass per particle of model galaxies
($\mpart$) is given in Table 1.

\begin{table}
 \caption{Parameters of the simulations}
 \begin{tabular}{@{}lccccc@{}}\\
 \hline
 %\hline
  Name     &   \mh(z=0)   &   $\mpart$   &  $\lmax$   &
$\fsf$  &   $\toff$  \\
           & ($10^{10}$ \msun)   &   ($10^{4}$ \msun)  &   (pc)  &   &
($10^{6}$ yr)  \\
 \hline
 
 $A_{h 1}$     & 1.97 & 9.41   & 109  & 0.5  & 20  \\ % B8 
$A_{h 1}^{*}$     & 2.01 & 9.41   & 109  & 0.5  & 20  \\ % B8* 
$B_{h 1}$        & 2.32 & 9.41   & 109  & 0.5  & 20  \\ % B9 
$B_{h 1}^{*} $    & 2.46 & 9.41   & 109  & 0.5  & 20  \\ % B9* 
$C_{l 1}$      & 2.80 & 65.7   & 218  & 0.5  & 20  \\ % C3 
$C_{h 1} $       & 2.48 & 8.28   & 109  & 0.5  & 20  \\ % C4 
$C_{h 1}^{*}$     & 2.84 & 8.28   & 109  & 0.5  & 20  \\ % C4* 
$D_{h 1} $       & 3.65 & 9.40   & 109  & 0.5  & 20  \\ % B7 
$E_{l 1}$      & 5.94 & 76.0   & 218  & 0.5  & 40  \\ % B1 
$E_{h 1} $       & 5.11 & 9.40   & 109  & 0.5  & 10  \\ % B4 
$E_{h 2} $       & 5.31 & 9.40   & 109  & 0.5  & 40  \\ % B2 
$E_{h 3}  $      & 5.76 & 9.40   & 109  & 0.2  & 40  \\ % B3 
$F_{l 1}$      & 7.79 & 65.7   & 218  & 0.5  & 20  \\ % C1 
$F_{h 1} $       & 6.56 & 8.28   & 109  & 0.5  & 20  \\ % C2 
$G_{l 1} $     & 8.37 & 76.0   & 218  & 0.66 & 40  \\ % A2 
$G_{l 2} $     & 9.52 & 76.0   & 218  & ---  & 40  \\ % A1 
$H_{l 1} $     & 14.5 & 46.5   & 218  & stoch.\tablenote{Stochastic SF scheme}  & r.a.\tablenote{"Run away" model, without turning off cooling}  \\ % D1 
$I_{l 1}  $    & 30.4 & 76.0   & 218  & 0.5  & 40  \\ % B5 
$I_{h 1}  $       & 30.5 & 9.40   & 109  & 0.5  & 40  \\ % B6 
$J_{l 1}  $     & 35.8 & 76.0   & 218  & 0.66 & 40  \\ % A4 
$J_{l 2}  $     & 40.4 & 76.0   & 218  & ---  & 40  \\ % A3 

\hline
\end{tabular}
\label{runs}
\end{table}

%=============================
\begin{figure}
%\centering
%\includegraphics[height=14cm,width=8.5cm]{f1.ps}
\vspace{12.4cm}
\includegraphics{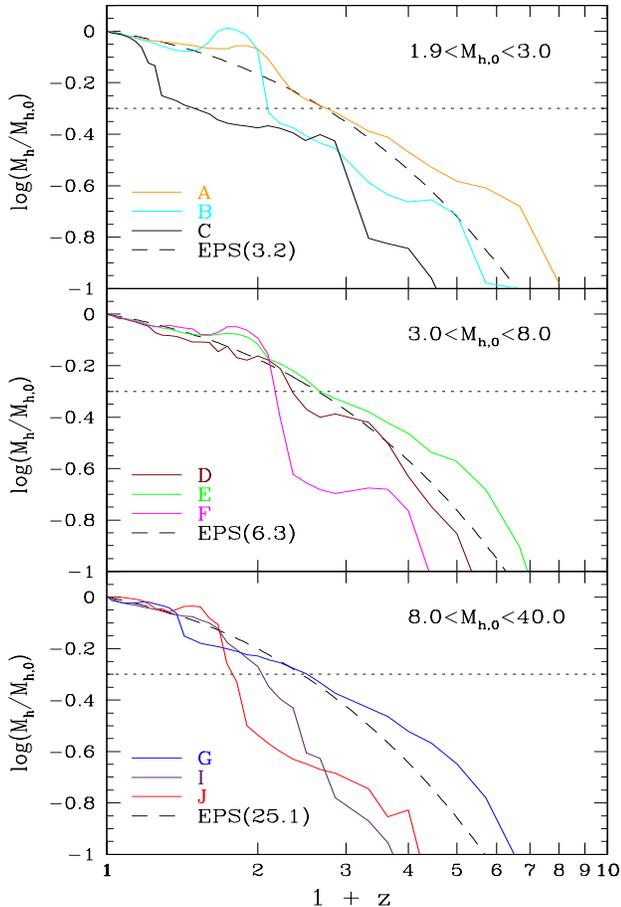}
\caption{Mass aggregation histories of  nine of our simulated halos normalized
to the mass attained at $z=0$, $M_{\rm h,0}$. In each panel, the MAHs of three halos
(identified with the uppercase letter) in the mass range shown in the upper-right parts 
of the panels are plotted ($M_{\rm h,0}$ is in unities of $10^{10}$ \msun).  The 
dashed curves correspond to the average MAHs (20,000 realizations) of halos of 
$M_{\rm h,0}$ as indicated in the parenthesis.
These MAHs were calculated using the EPS formalism by Firmani \& Avila-Reese (2000).
The horizontal dotted lines correspond to half $M_{\rm h,0}$.}
\label{MAHs}
\end{figure}
%=============================

In ART, the initially uniform grid is refined recursively as the matter
distribution evolves. The criterion chosen for refinement is based on gas or
DM densities. The cell is refined when the mass in DM particles 
exceeds 1.3$(1 - \fbU) m_p$ or the mass in gas is higher than 
13.0$\fbU\ m_p$, where $m_p = m_{dm} + m_g = m_{dm}/(1-\fbU)$ and \fbU\ is the
universal baryon fraction, \fbU = 0.15 for the cosmology used here;
it is assumed that the mean DM and gas densities in the box are the corresponding universal averages.
For the simulations presented in this paper, using multiple DM
particle masses, the grid is always unconditionally
refined to the third level (fourth level), corresponding to an 
effective grid size of $512^3$ ($1024^3$).  
On the other hand, the maximum allowed refinement level 
was set to 9 and 10, for low and high resolution simulations, 
respectively. This implies spatial sizes of the finest grid cells
of $\lmax=218$ and 109 comoving pc, respectively.

We have re-simulated, as described, the evolution of ten individual galaxy/halo systems covering 
a total halo mass (baryon + dark matter) range from  $\mh\approx 2 \times 10^{10}\ \msun$ 
to  $\mh\approx 4 \times 10^{11}\ \msun$ at $z=0$. For most of these systems,
we have run two or more simulations by varying the resolution, the SF local
efficiency parameter \fsf, the time during which cooling time (after
SF events) is kept off, $\toff$, and, in one case,
the SF scheme. Table \ref{runs} summarizes the characteristics of all
the simulations studied here. 

The first column gives the name of the run, denoted by a capital letter that refers 
to a particular galaxy/halo system (from $A$ to $J$ for ten systems), 
followed by a lowercase letter that is labeled ``$l$'' or ``$h$'' meaning 
low (218 pc) and high (109 pc) resolution,
respectively, and finally by a number that is used to differentiate cases with
different combinations of the parameters \fsf\ and $\toff$ (indicated in the next columns).
There are three simulations marked with the symbol ``*''. In simulations
with ``$h$'' but without ``*'', cells were refined if they satisfy the above mentioned condition
{\it but} with $m_p$ given by its {\it lower resolution} (less species) value, which is 8 times
larger.
Thus, the refinement condition for these simulations poorly obeys the cell
scaling and the ``$h$'' simulations end up with a number of resolution elements 
close to that obtained actually for ``$l"$ simulations. Model galaxies with ``$h''$ and ``*''
do satisfy the scaling: cells are refined according to the above recipe
with $m_p$ given by its higher resolution value. In other words,
simulations with ``$h$'' and ``*'' are of high resolution in the sense of the 
number of DM particles and in terms of number of grid cells.
As can be anticipated, the modeling of galaxies with ``$h$'' and ``*'' consumes 
a lot of CPU and wall time and thus, although desirable, not all models could 
be run with this resolution. In columns  (2), (3), and (4) the total halo virial mass \mh\ at $z=0$,
the DM particle mass resolution, $m_{dm}$, and the size of the finest grid cell, $\lmax$, are reported. 
The values used for the parameters \fsf\ and $\toff$ are shown
in columns (5) and (6). 

Galaxies $G$ and $J$ are actually from one simulation, corresponding to the box 
used in C10. This simulation was run to get a maximum
resolution of $\lmax=218$ pc (three DM particle species). 
Cases in which the \fsf\ column is written ``---'' refer to the simulations where $\fsf$
was not fixed but was calculated at each cell and timestep by an algorithm 
introducing other parameters (see \S\S 2.1 and C10). The system $E$ is the 
one that covers the largest variation of cases and parameters.

Galaxy $H$ was simulated with the stochastic approach for SF, $\epsilon_{SF}$ not 
fixed a priori, and without turning off the cooling. In this simulation, 
following \citet{CK2009}, "runaway stars"
are modeled. These are massive stars with high peculiar velocities as a result of the 
SN explosion in a close binary or due to two-body encounters in stellar cluster. 
As a result, runaway stars can move far away from the SF regions,
being then able to inject energy into the ISM in regions of low-density gas, 
where the cooling time is large so the feedback efficiency is higher.
The runaway stars are modeled through the addition of a random velocity, 
drawn from an exponential distribution with a characteristic scale of $17 \kms$, 
to the inherited gas velocity in 30\% of the stellar particles.

\subsection{Halo mass aggregation histories}

The halo mass aggregation histories (MAHs) of the simulations presented in Table 1, except for
run $H$, are plotted in Fig. \ref{MAHs}. The SF histories of galaxies are
expected to follow in a first approximation the halo MAHs.  As seen, the MAHs of the
different simulated distinct halos are quite diverse, and several of them oscillate around the
average \lcdm\ halo MAHs expected for these masses (dashed curves calculated with the 
Extended Press-Schechter -EPS- formalism of Firmani \& Avila-Reese 2000,
which gives results in good agreement with the fits to the outcomes of the Millenium
and Millenium II simulations (Fakhouri, Ma \& Boylan-Kolchin 2010)\footnote{
While the EPS average MAHs are for pure DM halos, the MAHs from the simulations
are for the DM+baryons systems. The inclusion of baryonic physics actually does not
affect significantly the halo MAHs.}. 
Most of our simulated halos show late (since $z\sim 0.6$) MAHs close to the 
corresponding averages.

Halos $B, F$, and $J$ suffer a late major merger 
at $z \approx 1.0$, 1.1, and 0.7, respectively. Actually, halo $F$ suffers two 
major mergers that occur around $z=1$; this explains the big jump
in mass seen in the MAH of this halo. Halos $A$ and $E$ have MAHs that
imply very early mass assembly.  Halos $C, F, I$ and $J$ have MAHs
implying mass assemblies later than the corresponding averages. In particular, is 
remarkable the delayed MAH of halo $C$. The increase in mass seen at  
$z \approx 0.33$ in this halo is due to the almost simultaneous accretion of 
three medium-sized halos each of mass of about one tenth of the mass of halo $C$, 
two of them coming in a pair. 

It is important to remark that all the simulated halos are distinct at $z=0$,
and the galaxies studied here are the {\it central ones} inside these halos.

\begin{table*} 
\begin{center}
\caption{Physical properties at $z=0$ of simulated galaxies} 
  \begin{tabular}{@{}ccccccccc@{}}\\ 
  \hline 
  \hline 
   Name     &   \mh\   &   \ms\   &  \fg\   &   \re\  &   \vmax\  &  SFR   &   \fs\   &  \fb\ \\ 
      (1)         & (2)   &   (3)  & (4)    & (5)  &  (6) &   (7) & (8)  & (9)  \\ 
            & ($10^{10}$ \msun)   &   ($10^{9}$ \msun)  &    & (kpc/h)  &  (km s$^{-1}$) &   (\msun\ yr$^{-1}$) &   &   \\ 
 \hline 
$A_{h 1}$        & 1.97 & 0.17   & 0.051  & 0.45  & 50.8  & 0.000  & 0.009  & 0.009 \\ % B8 
$A_{h 1}^{*}$     & 2.01 & 0.30   & 0.058  & 0.31  & 52.9  & 0.000  & 0.015  & 0.016 \\ % B8* 
$B_{h 1}$        & 2.32 & 0.22   & 0.440  & 0.59  & 49.4  & 0.003  & 0.009  & 0.017 \\ % B9 
$B_{h 1}^{*} $    & 2.46 & 0.21   & 0.690  & 0.50  & 55.9  & 0.009  & 0.009  & 0.027 \\ % B9* 
$C_{l1}$      & 2.80 & 0.10   & 0.850  & 0.66  & 57.1  & 0.000  & 0.004  & 0.023 \\ % C3 
$C_{h 1} $       & 2.48 & 0.21   & 0.480  & 0.94  & 44.2  & 0.006  & 0.009  & 0.016 \\ % C4 
$C_{h 1}^{*}$     & 2.84 & 0.16   & 0.760  & 0.74  & 51.4  & 0.012  & 0.006  & 0.024 \\ % C4* 
$D_{h 1} $       & 3.65 & 0.44   & 0.620  & 0.66  & 60.9  & 0.013  & 0.012  &  0.032\\ % B7 
$E_{l1}$      & 5.94 & 3.47   & 0.100  & 0.55  & 126.0  & 0.110  & 0.058  & 0.065 \\ % B1 
$E_{h 1} $       & 5.11 & 1.84   & 0.048  & 0.83  & 82.2  & 0.016  & 0.036  &  0.038\\ % B4 
$E_{h 2} $       & 5.31 & 2.02   & 0.340  & 1.05  & 84.0  & 0.044  & 0.038  &  0.058\\ % B2 
$E_{h 3}  $      & 5.76 & 3.56   & 0.280  & 0.80  & 104.8  & 0.062  & 0.062  & 0.086 \\ % B3 
$F_{l1}$      & 7.79 & 5.01   & 0.350  & 0.84  & 124.1  & 0.360  & 0.064  & 0.099 \\ % C1 
$F_{h 1} $       & 6.56 & 3.74   & 0.072  & 0.97  & 87.5  & 0.038  & 0.057  &   0.061\\ % C2 
$G_{l1} $     & 8.37 & 3.88   & 0.110  & 0.67 & 125.9  & 0.120  & 0.046  &  0.052 \\ % A2 
$G_{l2} $     & 9.52 & 7.82   & 0.220  & 0.55  & 185.0  & 0.650  & 0.082  & 0.106  \\ % A1 
$H_{l1} $     & 14.5 & 10.0   & 0.054  & 1.03  & 188.0  & 0.120  & 0.069  & 0.073 \\ % D1 
$I_{l1}  $    & 30.4 & 22.8   & 0.170  & 1.10  & 254.8  & 0.480  & 0.075  &  0.090\\ % B5 
$I_{h 1}  $       & 30.5 & 25.4   & 0.230  & 1.75  & 213.3  & 0.150  & 0.083  &  0.108\\ % B6 
$J_{l1}  $     & 35.8 & 20.9   & 0.070  & 1.21 & 214.3  & 0.150  & 0.058  &  0.063\\ % A4 
$J_{l2}  $     & 40.4 & 29.2   & 0.140  & 0.85  & 279.2  & 0.370  & 0.072  &  0.084\\ % A3 

\hline 
\end{tabular}  
\label{prop}
\end{center}
\end{table*}

%=====================
\section{Results and comparison with observations} \label{sec:results}
%=====================

Most of the simulated galaxies studied here show dynamical, structural, and ISM 
properties in reasonable agreement with observations of sub-\mstar\ central galaxies. The code 
SF/feedback schemes and their parameters (see \S 2.1) were extensively explored 
in C10. Based on the analysis results from that work, here we limit our experiments 
to only a few variations of some of the parameters and to simulations with 
higher resolutions (see Table \ref{runs}). 
Our goal is to explore the SFR, \ms\ and \mh\ evolution of low-mass isolated galaxies 
in the numerical simulations and to compare them to current observational inferences 
of central sub-\mstar\ galaxies at different redshifts.

\subsection{General properties}

The main properties at $z=0$ of all runs are reported in Table \ref{prop}.
The simulation name and halo virial mass \mh\ are given in columns (1) and (2).
Columns (3) and (4) present the galaxy stellar (disk + spheroid) mass and 
the (cold) galaxy gas mass fraction, \ms\ and \fg=\mg/(\ms + \mg), respectively, where 
\mg\ is the mass contained in gas particles with $T\le 10^4$ K.  
Both \ms\ and \mg\ are defined as the
stellar and gas mass contained within 5\re, where \re\ is in turn defined as the 
radius where half of the galaxy stellar mass is contained (column 5). This latter
mass is found inside the radius defined as the minimum between the tidal and
0.5 the halo virial radius. This definition considers the stellar mass
associated with a possible very extended stellar halo; satellite galaxies
would be also considered, but this makes no difference as far as \re\ and related 
properties are concerned because their contribution to the total mass is \lesssim\ 5\%.
The maximum circular velocity, \vmax, is given in column (6); 
the circular velocity is computed as $V_{\rm c} = \sqrt{GM(R)/R}$, where $M(R)$
is the total (dark, stellar, and gas) mass. Finally,
columns (7) and (8)  give the galaxy stellar and baryon mass fractions,  
defined as \fs=\ms/\mh\ and \fb=\mb/\mh, where \mb=\ms + \mg\ and
\mh\ is the total (dark+baryon) virial mass. 

\begin{figure}
%\plotone{f2.ps}
\vspace{9.5cm}
\includegraphics{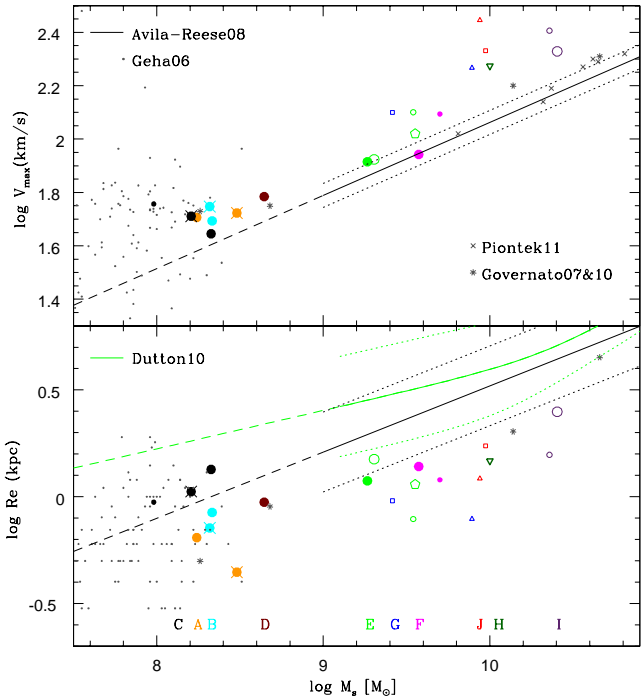}
\caption{Maximum circular velocity (upper panel) and effective radius (lower panel)
vs. \ms\ for the different galaxies simulated here (figure symbols, see text
for the symbol code and Table 1 for the color/letter code showed below in lower panel). 
The small skeletal symbols are for
the simulations from \citet{Piontek09} and Governato et al. (2007, 2010).
The fits and 1$\sigma$ intrinsic scatters in both panels to a compiled sample of normal 
disk galaxies by \citet{Avila08} are plotted with solid and dotted lines, respectively.
The observed dwarf galaxies by Geha et al. (2006) are also plotted (dots).
In lower panel, the fit to the blue-cloud disk-dominated galaxies from the SDSS 
by \citet{Dutton10c} is reproduced (green solid line; the dotted lines 
show the 84th and 16th percentiles of the size distribution).  The dashed
lines are extrapolations to lower masses of the given fits.}
\label{scale-rel}
\end{figure}

In Figure \ref{scale-rel}, \vmax\ vs \ms\ (the stellar Tully-Fisher relation,
sTFR, panel a) and \re\ vs \ms\ (panel b) at $z=0$ are plotted for all 
simulated galaxies (geometric symbols). The black solid and dotted lines 
show linear fits and their estimated $1\sigma$ intrinsic scatters to observed normal 
disk galaxies in a large mass and surface density range as reported in \citet{Avila08}. 
Their stellar masses were corrected by $-0.1$ dex to go from the 'diet Salpeter'
IMF \citep{Bell03} implicit in their inferences to the \citet{Chabrier03} IMF.  

Big and small geometric symbols are used for high- and low- resolution runs,
respectively. The symbols are open when $\toff=40$ Myr and solid
when $\toff=20$ or 10 Myr. Circle, pentagon, and square are used
for runs where \fsf= 0.5, 0.2, and 0.67, respectively; a triangle is used
when the stellar mass particle is calculated rather than assigned as 
a fraction of the gas mass cell (see \S\S 2.2).  The runs that correspond
to the same galaxy/halo system have the same color.  Symbols traversed 
by a cross correspond to high-resolution simulations with more grid cells 
($A_{h1}^{*}$, $B_{h1}^{*}$, and $C_{h1}^{*}$). The open inverted 
triangle corresponds to the simulation with the run-away stars scheme 
and without turning off cooling.

%===================================================
%===================================================

From upper panel of Fig. \ref{scale-rel}, one clearly sees that the low-resolution 
simulations produce galaxies with too large values of \vmax\ for their masses. 
It is well known that the lack of 
resolution in this kind of simulations introduces an artificial dissipation of 
energy and angular momentum in the gas, producing these too-concentrated 
galaxies and therefore, peaked rotation curves (as well as less SF and energy
input to the gas, in such a way that galaxy outflows are inefficient, having then
the galaxies too high baryonic and stellar fractions, see Table 2). However, our 
high-resolution simulations are already in rough agreement with the 
observationally-inferred sTFR, except for the run $I$. In general, the mid-plane 
gas disk rotation velocity profile, $V_{\rm rot}(R)$, is lower
than $V_{\rm c}(R)$ in the inner regions \citep[e.g.,][C10]{Valenzuela07},
and in some cases, even the maximum of $V_{\rm rot}(R)$ may be 
slightly lower than \vmax. Therefore, when comparing with observations,
\vmax\ tends to be an upper limit.
In Fig.  \ref{scale-rel} we also reproduce results from other recent high-resolution
simulations \citep[][]{Governato07,Governato10,Piontek09}, 
where the problem of obtaining a TFR shifted to the high-velocity side
seems to have been partially overcome (see \S\S4.1 for details on these works).

Our smallest simulated galaxies fall systematically above the extrapolated 
sTFR of higher masses, suggesting a bend towards the low 
\ms\ (high \vmax) side. Such a bend at low masses in the sTFR has been 
discussed recently by \citet{derossi10}, who explain it as a consequence 
of the strong stellar feedback-driven outflows. 
From the observational side, some authors have found that the infrared
or stellar TFRs at low masses become very scattered but apparently 
\ms\ is on average lower for a given \vmax\ than the extrapolation of the 
high-mass sTFR \citep[e.g.,][]{Mcgaugh05,Geha06,deRijcke07}. 
In the upper panel of Fig. \ref{scale-rel}, the observational data for dwarf
galaxies reported in \citet{Geha06} are reproduced (dots).

Regarding the \re--\ms\ relation, the lower panel of Fig. \ref{scale-rel} shows
that the low-resolution simulations produce too concentrated galaxies,
in agreement with the resolution arguments already mentioned above. 
However, even for the high-resolution simulations, our most massive modeled 
galaxies have radii smaller than the mean of the observational inferences for a 
given \ms. In Fig. \ref{scale-rel} is also plotted the fit to the  
\re--\ms\ relation inferred recently by \citet{Dutton10c} for the blue-cloud 
disk-dominated galaxies from the SDSS (green solid line). 
These inferences show a pronounced bend (flattening) in the \re--\ms\ 
relation at low masses ($\ms\lesssim 2\ 10^{10}$ \msun). The measured radii
for dwarf galaxies from \citet{Geha06} are also reproduced in Fig. \ref{scale-rel}
(dots). These galaxies lie clearly below the extrapolation of the the \citet{Dutton10c}
\re--\ms\ relation, but agree with the extrapolation of the \citet{Avila08} relation.
Our lowest-mass simulations show better agreement with observations.

It should be noted that \re\ both in \citet{Avila08} and \citet{Dutton10c} was 
calculated assuming $\re = 1.67 R_d$, where $R_d$ is the
scale radius obtained from fitting an exponential law to the disk 
component of the surface brightness profile. The above relation is
a good approximation if most of the stellar mass lies in a disk; however,
if the bulge/spheroidal 
component is non-negligible, then the actual effective radius \re\ 
is expected to be different from $1.67 R_d$.  For several of our high-resolution 
simulations, we have found that  \re\ is smaller than $1.67 R_d$ by 
factors of $\sim 1.1-1.3$, not enough to explain the differences seen in
Fig. \ref{scale-rel} for the more massive galaxies. In some
cases, \re\ results even larger than $1.67 R_d$. 

The structure of the stellar component of normal low-mass and dwarf galaxies 
seems to be different from the usual one of normal high-mass galaxies, both from 
the point of view of observations and simulations. In particular, the existence of 
an extended spheroidal component could be the rule as seen in our lowest-mass simulations 
\citep[see also C10;][and therein references on observational works]{Bekki08}.  
In general, the stellar surface density profiles of our simulated galaxies tend 
to be exponential within $\sim 1-3$ \re.

\begin{figure}
%\plotone{vel_m_all.ps}
\vspace{6.4 cm}
\includegraphics{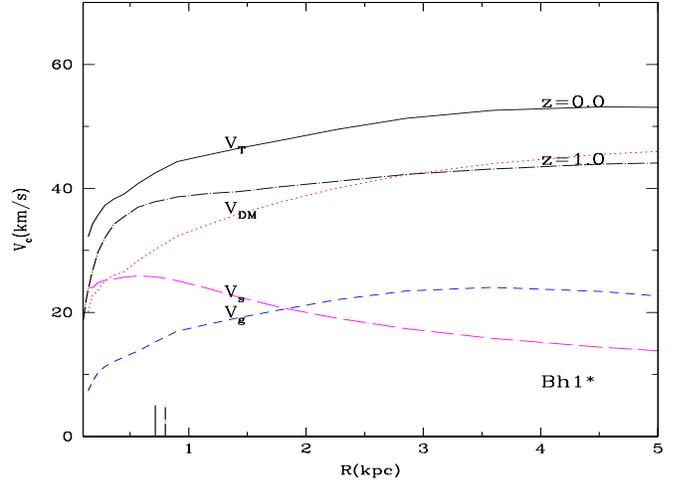}
\caption{Circular velocity profile, $V_{\rm c}(R)$, at $z=1$ (thick dot-dashed line)
and $z=0$ (solid line) of run $B_{h1}^*$. The velocity components at $z=0$ 
corresponding to dark matter (dotted red line), galaxy stars (long-dashed pink line), 
and galaxy gas (short-dashed blue line) are also plotted. The vertical segments
above the x-axis indicate $1 \re$ at $z=1$ and $z=0$.
\label{Vevol}}
\end{figure}

Regarding the circular velocity profiles, $V_{\rm c}(R)$, they are roughly flat at 
$1.5-3$ \re\ for all the high-resolution runs except for the most massive one, $I_{h1}$, 
which shows a narrow peak at 0.2\re. In some cases, $V_{\rm c}(R)$ is even 
slightly increasing at $1-2$ \re\ as it is the case of run 
$B_{h1}^*$, shown at $z=0.0$ (solid line) and 1.0 (thick dot-dashed line)
in Figure \ref{Vevol}. For our lower-resolution simulations, in some cases
the velocity profiles are peaked. 
As expected, $V_{\rm c}(R)$ increases with time. 
The decomposition of $V_{\rm c}(R)$ into the dark matter (DM), stellar (s) and
gas (g) components is also plotted in Fig. \ref{Vevol}, for the $z=0$ 
profile. This galaxy is dark-matter-dominated
for radii $\gtrsim 1 \re$.  The effect of stellar 
feedback is crucial in order to prevent excessive baryon matter concentration 
in the center --and consequently a peaked $V_{\rm c}(R)$ profile-- as has been 
discussed in C10 \citep[see also][]{Governato07,CK2009,Stinson09}. 

%===================================================
%===================================================

%===================================================
\begin{figure}
%\plotone{f4.ps}
\vspace{8.cm}
\includegraphics{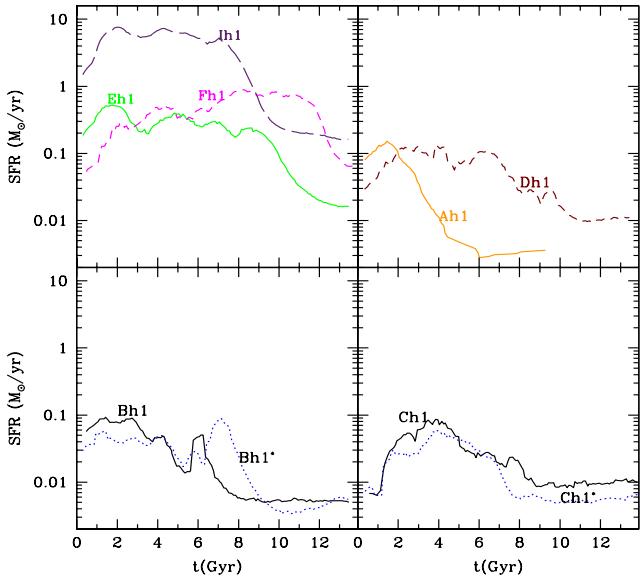}
\caption{"Archaelogical" SFHs from the high-resolution runs. Top-panels:
$I_{h1}, F_{h1}, E_{h1}$ runs (left panel) and $D_{h1}, A_{h1}$ runs (right panel).
Lower-panels: A comparison of the $B$ run with and without ``*''; that is,
run with a low and high number of cells inside \rv\ (left panel). Right-panel:
Same as in left panel but for run $C$. With the exception of run $F_{h1}$, models show
a much smaller SFR in the last 5-6 or more Gyr with respect to the previous,
active, early epoch.
\label{SFRHs}}
\end{figure}
%===================================================

Figure \ref{SFRHs} presents the ``archeological'' SF histories (SFHs) 
for our high-resolution runs, defined as the instantaneous SFR as a function 
of time, and computed for each run using the last snapshot recorded. Specifically, 
in each data dump, in addition to the positions and velocities for all stellar
particles, the code also saves the time at which they formed, their masses (initial
and present), and metallicities due to SNe of types Ia and II.  
We add up the (initial) masses of the stellar particles
formed during a certain time interval, which we take as 0.5 Gyr, and
divide it by this time, to obtain the ``instantaneous'' SFR. 
Since the amount of stellar mass outside the
galaxy is typically a small fraction, the SFH computed
inside \rv\ is mostly that of the central galaxy.
Unlike C10, where a time interval of 0.1 Gyr was taken to 
compute the SFHs, here we use 0.5 Gyr in order to produce smoother SFHs. 
%The bursty nature of models, highlighted by a smaller time bin, 
%is similar to that found in C10 for the ``deterministic'' SFR prescription. 
To estimate how much the SFR fluctuates, for
example, in the last 1 Gyr, we measured the SFR in bins of 0.1 Gyr width 
and computed the dispersion of data for each run. We find dispersions 
around the means of about 30-40\%. 

In the top panels, the ``archeological'' SFHs of runs $I_{h1}, 
F_{h1}, E_{h1}$ (left panel), and $D_{h1}, A_{h1}$ (right panel), 
are plotted, using the same color coding as in Figure 1. 
In the lower-left panel, we compare the SFHs of run $B$ simulated
with a high number of DM matter particles but with a 
low (dotted blue lines) and high (solid black lines) number of
cells. The same comparison but for run $C$ is shown in the lower-right
panel. As can be seen, there is not a significant difference 
in their SFHs between models with ``*'' and those without it.
The SFHs are diverse but in most cases they are characterized
by an active phase that starts after the first 1-2 Gyr and
lasts 2-6 Gyr. This phase is followed by a ``quiescent'' stage with
a low value of the SFR, with most of the SF concentrated in 
the central part of galaxies. In most cases, the SFHs inherit some
features of the halo MAHs. The ``archaeological'' 
SFHs already forecast that simulated low-mass galaxies will have 
low SSFRs at late epochs in conflict with observational inferences.

Finally, it should be said that in almost all of our simulations
a multi-phase ISM develops in the disks (see for more
details C10). This results from the efficiency by which thermal energy
is injected into the ISM and the ability of the medium to self-regulate
its SF.

\subsection{Specific star formation rate as a function of mass} 

\begin{figure*}
%\plotone{f5.ps}
\vspace{7cm}
\includegraphics{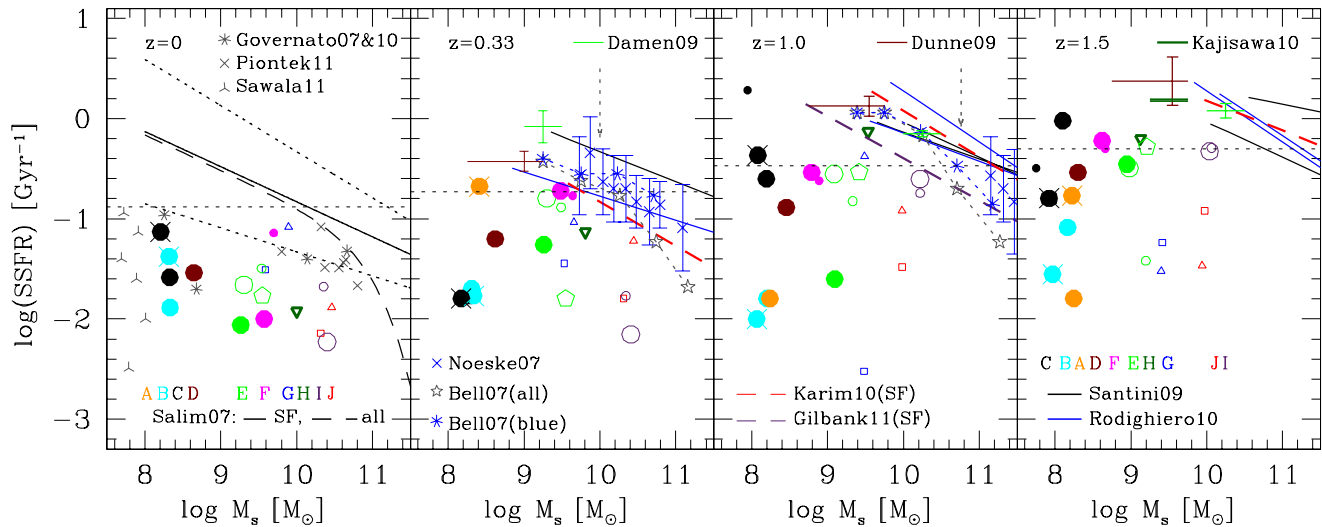}
\caption{SSFR vs \ms\ measured at four different redshifts for all of our
simulations (geometrical symbols). A given galaxy is represented with the 
same color (associated to the letter that identifies the simulation according 
to Table 1), and the different symbols are to distinguish runs of the same 
galaxy but with different sub-grid parameters (see \S\S .. for the symbol code).  
The horizontal dotted lines in each panel correspond to the SSFR of a galaxy
that had a constant SFR in the past (see text); galaxies above (below) this
line are expected to have been less (more) active in forming stars on average 
in the  past than in the current epoch. 
In the $z=0$ panel results from simulations by Governato et al. (2007, 2010),
\citet{Piontek09}, and \citet{Sawala10} are also plotted; the
solid line is the linear fit carried out by \citet{Salim07} to the sub-sample
of star-forming galaxies from a volume-corrected sample of $\sim 50,000$ 
SDSS galaxies (the intrinsic scatter is shown with dotted
lines), while the long-dashed line is the Schechter fit to the entire sample,
i.e. including composite galaxies (SF/AGN), galaxies dominated by AGN
emission, and those with no H$\alpha$ detection. The skeletal symbols, 
fit lines, and horizontal segments in the high-redshift panels 
correspond to different observational estimates down to the completeness limits
reported in the works indicated inside the panels (see Appendix for a 
description of each one of them).  
At all the epochs reported here, the observationally estimated SSFRs of low mass-galaxies 
are on average significantly higher than those of simulations. While 
observational estimates at each one of the redshifts are typically above the
SFR=const. lines for low masses, the simulations are in most cases
below them, specially for $z<1$.}
\label{SSFR}
\end{figure*}

The main result of this paper is shown in Figure \ref{SSFR}, 
where the {\it directly} measured galaxy SSFR of the simulations at four epochs
($z=0.00, 0.33, 1.00$ and 1.50) are plotted vs the corresponding current stellar 
masses (geometric symbols). For the symbol code and color/letter code,
see previous subsection. Unlike the ``archeological'' SFR shown in
Fig. \ref{SFRHs}, here the SFR at a given redshift is computed by
summing up the mass of all stellar particles with ages smaller than 0.1 Gyr
located in a cylinder with 1.0 \kpch\ (proper) height and 20.0 \kpch\
of radius, centered on the gaseous disk. The SFR is then this mass
divided by 0.1 Gyr. This time bin is 
wide enough so as to avoid short fluctuations in the SFR values. 

The general trend of the measured SSFR in the simulations is to decrease with time,
though, in general the SFR (and therefore SSFR) histories are quite episodic  
(see also Fig. \ref{SFRHs}).
It is remarkable that at low redshifts ($z=0$ and 0.33), all simulated galaxies 
have SSFRs below the SSFR a galaxy would have if it had formed all
stars at a constant SFR\footnote{If SFR=const. in time, then 
SFR/\ms = 1/[(1 -- R)($t_H$($z$) -- 1 Gyr)], where $R=0.4$ is the 
average gas return factor due to stellar mass loss, $t_H$ is the cosmic time, and 
1 Gyr is subtracted in order to take into account the onset of galaxy formation.}
(horizontal dashed line). 
This means that the \ms\ of these galaxies was assembled relatively early,
the current SFRs being smaller than their past average SFRs (the galaxies entered 
into a quiescent SF phase). For higher redshifts, $z=1.00$ and 1.50, the SSFRs 
of several of the {\it high-resolution} simulations tend to be already around the SFR=const. line, 
which means that galaxies are in their active phase of SF and \ms\ assembly.

Notice that some runs show negligible SSFRs at some epochs. For example, the 
SSFR of run $A_{h1}^*$ is zero at $z = 0.0$ (the halo MAH of this run 
is the earliest one among all simulations, see Fig. \ref{MAHs}, which
seems to imply that its SFR decreased strongly very early, see Fig. \ref{SFRHs}). 
Because of the episodic nature of the SFR it may 
well be that, even though we use a relatively wide time bin to compute the 
instantaneous SFR (0.1 Gyr), there are some epochs for which we see a 
galaxy almost not forming stars. Run $C_{l1}$ is special in the sense that it has
zero SFR at the time-steps corresponding to $z=0$ and $z=0.33$, something 
that is not seen for the higher-resolution counterpart simulations. 
%%%%%  decidir si incluir lo que sigue %%%%%%%
The resolution
in this case seems to affect the efficiency of SF because the galaxy actually is 
growing plenty of gas ($\fg = 0.85$ and 0.70 at $z=0$ and 0.33, respectively; note
that the halo MAH of this model shows the most active growth at late epochs).
This is probably because instabilities in the disk have not been resolved. Indeed, 
in the corresponding high-resolution runs, specially the $C_{h1}^*$ one,
a relative high SSFR is measured at late epochs. However, there are cases
of high-resolution simulations, for example run $D_{h1}$, where the gas fraction 
is high and even increases at late epochs, but the SSFR keeps low and
decreasing with time.  We suspect that in these cases some physical 
process is damping the disk gas instabilities, and therefore inhibiting active SF; the 
presence of a dynamically hot spheroid --which is actually formed
in several of our low-mass simulations-- has been shown to work
in this direction  \citep[][]{Martig09}.

%=============================================
%=============================================

From Fig. \ref{SSFR} one sees that the resolution affects moderately
the SSFRs. Our low-resolution simulations tend to have lower values
of SSFR at $z=1.50$ and 1.00 than the high-resolution ones, but
the opposite applies for lower $z$. At this point, it should be said  that 
it is difficult to establish any systematicity due to the episodic behavior of 
the SFR in the simulations. At $z=0$, low-resolution simulations seem to 
produce slightly higher SSFR values than high-resolution
simulations. Note also that the former have systematically higher values
of \fg\ than the latter (see Table 2). The effect of increasing the number 
of cells (filled circles traversed by a cross) is 
not so significant.  The global effect of the strength of the stellar feedback 
(regulated in our simulations by $\fsf$ and $\toff$) 
apparently is to keep the SF more active at
later epochs as seen for runs $E_{h1}$ ($\toff = 10$ Myr, open green circle) and 
$E_{h2}$ ($\toff$= 40 Myr, filled green circle). However, the differences are
small, as in the case of variations in \fsf\ from 0.5 to 0.2 (runs $E_{h2}$, filled 
green circle, and $E_{h3}$, filled green pentagon, respectively).
In general, variations of the sub-grid parameters around reasonable values 
seem not to be a source of significant and systematical changes in the SSFR
of the simulated galaxies.

\subsubsection{Comparison with observational inferences}

In each panel of Fig. \ref{SSFR}, different observational inferences compiled
here are plotted for comparison. Unless otherwise stated, we plot only the
data that obey  the completeness limit given by the different authors. 
The inferences of \ms\ (and SFR) from the 
observed luminosities or spectral energy distribution
of galaxies are based on results from SPS (stellar population synthesis) 
models. For all
cases reproduced in Fig. \ref{SSFR}, a constant, universal, stellar 
initial mass function (IMF) has been assumed in these models, although
different authors may have used different IMFs. We homogenize all 
observational results to a \citet{Chabrier03} IMF (see Appendix for
the different corrections applied here).  A description of each one
of the observational sources compiled from the literature is presented in 
the Appendix. 

%====================================================
\begin{figure*}
%\plotone{f6.ps}
\vspace{7cm}
\includegraphics{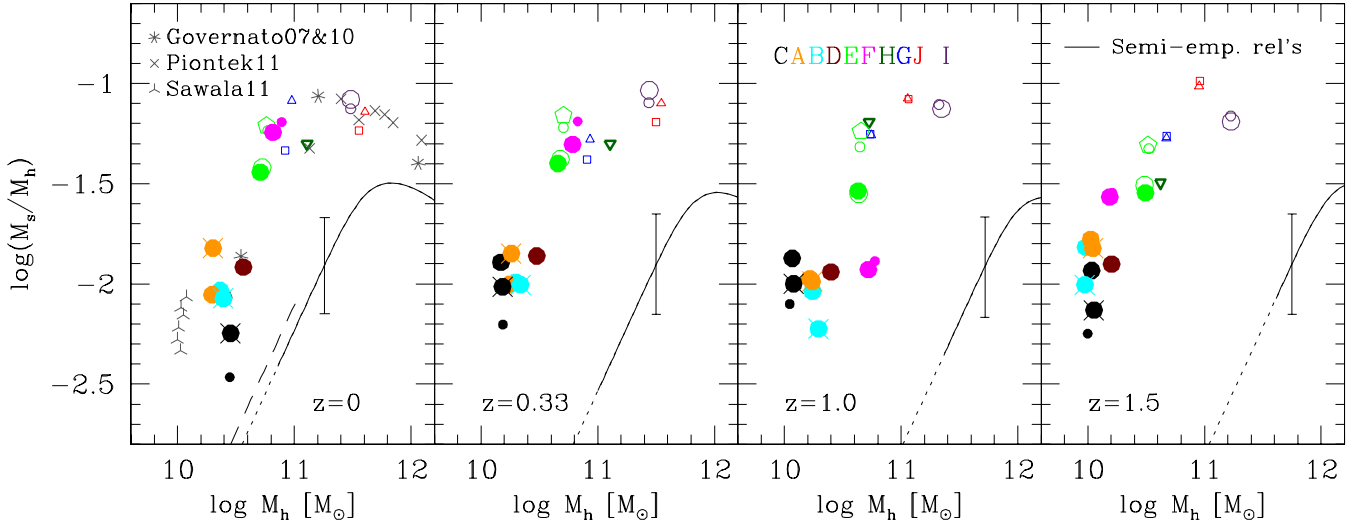}
\caption{Stellar mass fraction vs \mh\ of the simulated galaxies at
$z=0,$ 0.33, 1, and 1.5 (geometrical symbols). The symbol and
color/letter codes are as in Fig.\ref{SSFR}. The results from other
authors (indicated inside the panel) are also reproduced for $z=0$.
The solid curves in each panel are semi-empirical inferences
as reported in FA10 based on \citet{BCW10}. The error bars show the
$1\sigma$ uncertainty of these inferences. The dotted lines are extrapolations
to lower masses, given the lower limits in \ms\ of the empirical galaxy \ms\ 
functions used in Behroozi et al. 
For the dashed line in the $z=0$ panel see the text. 
\label{fsMh}}
\end{figure*}
%====================================================

From a visual inspection of Fig. 
\ref{SSFR}  we conclude that all simulated low-mass galaxies at different 
resolutions and sub-grid parameters lie significantly below the averages
of a large body of SSFRs observational determinations as a function of 
\ms\ out to $z\sim 1.5$, especially for the blue star-forming ones.   
In the $z=0$ panel, a compilation of recent numerical results from other 
authors are also plotted (skeletal symbols; see \S\S 4.1 for more details).

By taking into account only our high-resolution simulations with 
$\fsf=0.5$, we estimate that the SSFRs in the mass
range $\ms\sim 10^8-10^{10}$ \msun\ are $\sim 5-10$ times
lower than the means inferred from observations at $z\approx 0$.
These differences are roughly the same or slightly smaller 
at $z\approx 0.33$. At higher redshifts ($z\approx 1.0-1.5$), the 
comparison becomes difficult by the incompleteness in the observational 
samples. However, both for the small mass range where our simulations
coincide with observations above the completeness limit and the extrapolations
to lower masses of the inferred SSFR--\ms\ relations,  
the differences persist, though they are less dramatic than at lower $z$.

It is notable the systematical shift of the SSFR--\ms\
relation of observed galaxies as $z$ decreases: from a rough average 
of all relations, the typical stellar mass, $M_{\rm quench}$, that crosses 
below the line corresponding to the constant SFR case at each $z$ 
(horizontal short-dashed lines in Fig. \ref{SSFR}) decreases with time. 
In other words, the lower  the $z$, the smaller on average are the galaxies 
that start quenching their SF (downsizing in SSFR).  Interesting enough, the 
downsizing of $M_{\rm quench}$ is in rough quantitative agreement 
with the downsizing of the transition mass from active to passive 
galaxies found from the semi-empirical \ms\ growth tracks in 
FA10: log$M_{\rm tran}$($z$) $= 10.30 + 0.55z$.  
{\it For the simulated galaxies, there is no evidence
of such a phenomenon of downsizing.}

\subsection{Stellar mass fraction as a function of mass}

In Fig. \ref{fsMh} the stellar mass fraction, $\fs\equiv \ms/\mh$, of our
simulated galaxies vs \mh\ at $z=0.00,$ 0.33, 1.00 and 1.50 are plotted.
Results from numerical simulations of other authors 
shown in Fig. \ref{SSFR} are also plotted (skeletal symbols).

The bell-shaped curve in each panel corresponds to the continuous
analytical approximations given in FA10 to the semi-empirical determinations of the \ms--\mh\
relations at different redshifts ($0\lesssim z< 4$) performed by \citet{BCW10}. These
inferences are based on the abundance matching formalism, where
the cumulative observed galaxy stellar mass function at a given epoch 
is matched to the cumulative halo mass function in order to find the
\mh\ corresponding to a given \ms, under the assumption of a
one to one galaxy-halo correspondence.  Several authors have found 
similar results out to $z\sim 1$, even when different observational data sets and 
formalisms have been used \citep[see for recent
results e.g.,][]{Moster10,Guo10,WangJing10}. For the local universe,
direct techniques, as galaxy-galaxy weak lensing and satellite kinematics, 
have also been applied to infer the \ms--\mh\ relation
with results that are in reasonable agreement among
them and with other techniques \citep[see for comparisons][and more
references therein]{BCW10,More10,R11} in the mass ranges where they
can be compared (the direct techniques are reliable for the time being  
only for intermediate or massive galaxies).  The error bar in Fig. \ref{fsMh}
indicates the approximate $\pm1\sigma$ statistical and sample variance
uncertainties in the inferences. The
main contributor to this uncertainty comes from assumptions
in converting galaxy luminosity into \ms\ \citep{BCW10}. 

According to Fig. \ref{fsMh}, the smaller is the mass, the lower is \fs\ on 
average, both for our simulations and the semi-empirical inferences. 
However, the simulation results have \fs\ values systematically higher 
than those inferred from observations. At $z\approx 0$, the differences amount to
factors around 5--10 and they apparently increase at higher redshifts;
at $z=1.5$, the simulations have values of \fs\ for a given \ms\ around 30 times 
higher than those inferred semi-empirically at the masses where the comparison
can be done; the same factor applies for the lowest masses if one extrapolates 
the semi-empirical determination to these masses. For simulated galaxies,  \fs\
changes very little in the redshift range reported here while
halo masses decrease as $z$ increases. Therefore one expects the 
\fs--\mh\ relation to shift on average to the low-mass side as $z$ gets higher. 
This behavior is contrary to the semi-empirical inferences.  

In the semi-empirical inferences, the mass function of pure dark matter
halos is used. It could be that the masses of the halos (dark+baryonic matter)
end up smaller when baryonic processes are included; for example, as a
result of mass loss out the virial radius in low-mass halos. The maximum fraction of 
ejected baryons at $z=0$ in our simulations is $82\%$ for runs $A_{h1}^*$ and
$B_{h1}$\footnote{For other mass runs with $\mh<3\times 10^9$ \msun, this fraction vary 
around 60-80\%, while for the more massive runs the fractions are $10-40\%$.}.
This would imply that \mh\ in run $A_{h1}^*$ (or $B_{h1}$) is $\approx (1- 0.82\fbU)^{-1} =1.14$ times
smaller than in a purely dark matter simulation. 
The dashed line in Fig. \ref{fsMh}, $z=0$ panel, shows the expected shift in the 
$\ms/\mh-\mh$ relation by this maximum factor.
On the other hand, the inferred \ms/\mh\ ratio at low masses depends mainly on 
what is the faint-end slope of the galaxy \ms\ function. If observations show that 
there is a steepening of this slope, then the \ms/\mh\ ratio may flatten at low masses,
being in better agreement with our simulation results. Some recent observational
studies suggest that the higher the $z$ (for $z\gtrsim 1$), the steeper is the faint-end slope
\citep{Kajisawa09,Mortlock11}. For lower $z$, there is also evidence of 
a somewhat steep slope \citep[$\sim -1.7$, e.g.][]{Drory+2009}. However,
even that, the \ms/\mh--\mh\ relation at low-masses seems to be steeper
and lower than in our simulations \citep[see Fig. 9 in][]{Drory+2009}. 

\begin{figure}
\vspace{7.cm}
\includegraphics{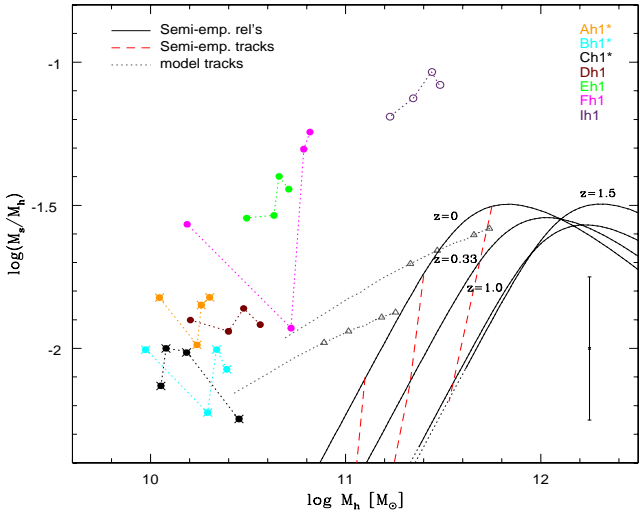}
\caption{Stellar mass fraction vs \mh\ at $z=0, 0.33, 1.0$ and 1.5 for the high-resolution
runs indicated inside the panel, and for two semi-numerical models presented in 
FA10 (open black triangles; the dotted lines extend up to $z=4$). The solid black 
curves are the same semi-empirical determinations plotted in Fig. \ref{fsMh} at the 
four epochs mentioned above. The dashed red curves correspond to
three semi-empirical individual {\it average} evolutionary tracks (GHETs) as 
inferred in FA10. 
\label{fsMh_evol}}
\end{figure}

In Fig. \ref{fsMh_evol}, we have merged into one the four panels (epochs) of
Fig. \ref{fsMh}, showing the evolution (connected with dotted lines) of only some of
our high-resolution simulations. The relatively large jumps seen in some runs
are mainly due to halo major mergers. For example, run $F_{\rm h1}$ shows a
strong decreasing of \fs=\ms/\mh\ at $z\approx 1$ because it suffered 
major mergers around this epoch (see Fig. \ref{MAHs}). For simulations
where the halo MAH tends to be smooth, \fs\ changes more
smoothly with $z$ (e.g., runs $D_{\rm h1}$, $E_{\rm h1}$ and $I_{\rm h1}$).

We also plot in Fig. \ref{fsMh_evol} the evolution of two semi-numerical 
models presented in FA10. These models 
correspond to the case of energy-driven SN outflows and no re-accretion,
with an outflow efficiency of 65\%; only with such a high efficiency it is possible
to lower the stellar mass fractions to roughly agree with the
\fs--\mh\ determinations at  $z=0$. 

For both cases, simulations and semi-numerical
models, \fs\ changes relatively little with $z$.  As mentioned in the Introduction, FA10
 \citep[see also][]{CW09} connected the \fs--\mh\ relations by using
 {\it average} MAHs and obtained the corresponding \textit{individual} (average) \fs\ (or \ms) 
hybrid evolutionary tracks (GHETs, dashed red curves in Fig. \ref{fsMh_evol}). These tracks in the 
log\ms($z$)(or log\fs)-log\mh($z$) plane have very steep slopes for low-mass galaxies, 
 $d$log\ms($z)$/$d$log\mh($z$)$\sim 4.5$ ($z<1$), i.e. for these galaxies, \ms\ (\fs) 
 grows very fast since $z\sim 1$. However, for the semi-numerically 
 modeled galaxies,  $d$log\ms($z)$/$d$log\mh($z$)$\sim 1.4$, 
i.e. \ms\ (\fs) grows slowly. {\it Our numerical simulations, albeit noisily, confirm such a 
behaviour, which strongly disagrees with the semi-empirical tracks. }

%===================================================
%===================================================

%===================================================
\section{Discussion}
%===================================================

We would like to stress that (i) the potential issues of low-mass \lcdm-based 
simulated galaxies showed in Figs. \ref{SSFR} and \ref{fsMh} seem to be shared 
by most previous high-resolution numerical simulation works, and (ii) that the empirical 
inferences we have compiled in this paper are yet subject to large 
uncertainties and probable systematical errors. 
Following, we discuss both items.

\subsection{Previous numerical works and numerical issues}

In Figs. \ref{scale-rel}, \ref{SSFR} and \ref{fsMh} we have reproduced the $z=0$ properties
of sub-\mstar\ galaxies presented in recent numerical works, where very high-resolution
re-simulations of individual low-mass galaxies were carried out and the
quantities we are interested in were reported.  

The simulations by  \citet{Governato07,Governato10} (asterisks 
in Figs. \ref{scale-rel}, \ref{SSFR}, and \ref{fsMh}) were performed
with GASOLINE, a N-body + Smoothed Particle Hydrodynamics (SPH) code. 
The SF prescription ensures that  the SFR density is a function of the 
gas density according to the observed slope of the Kennicutt-Schmidt law, 
and a SF efficiency parameter sets the normalization of this relation. In order 
to allow for SN-driven expanding hot bubbles in the non-resolved regions, 
cooling is turned off in the gas particles receiving SN energy until the end of the snowplow 
phase according to the Sedov-Taylor solution. 
For the force resolutions of $\sim 100-200$ pc achieved in their low-mass simulations,
\citet{Governato10} used a high value for the SF density threshold, $\nsf = 100\ \pcc$ \citep[for the
high-mass simulations in][$\nsf = 0.1\ \pcc$ was used]{Governato07}, 
in order to have enhanced gas outflows that remove low angular momentum
gas from the central regions of the galaxy; according to the authors, this helps 
to create realistic disks.

The simulations by \citet{Piontek09} (crosses in Figs. \ref{scale-rel}, 
\ref{SSFR}, and \ref{fsMh}) were performed with GADGET-2, a Tree-PM 
N-body + SPH code. The SFR was implemented according to a Schmidt law, 
$\dot{\rho}_* = c_*\rho_g/t_{SF}$, where $c_*$ is a SF efficiency
parameter, and  a stochastic approach for assigning stellar particles to the gas 
particles masses was used. Cooling is
turned off in the gas particles receiving SN energy by a fixed time, 20 Myr.
We reproduce here their "standard model" high-resolution simulations 
(gravitational softening parameters of $100-140$ pc). With few exceptions,
they find little or no angular momentum deficiency in their galaxies, though
prominent bulges persist in most cases and galaxies lie slightly
shifted in the stellar TFR. 

The simulations by  \citet{Sawala10} (skeletal triangles in Figs. \ref{SSFR} and \ref{fsMh})
correspond to six representative low-mass halos (\mh($z=0$)$\sim 10^{10}$ \msun) 
extracted from the Millenium-II Simulation. These halos were resimulated
at high resolution with gas (dark and stellar mass particles
of $8\times 10^4$ and $1.8\times 10^4$ \msun, respectively) using the Tree-PM
code GADGET-3, which includes metal-dependent cooling, SF (Schimdt
law and stochastic approach), chemical
enrichment, and energy injection from type II and Ia SNe implemented in a 
multiphase gas model developed in \citet{Scannapieco06}.  SN energy is 
shared equally between the hot and cold phases. Cold particles which
receive SN feedback accumulate energy until their thermodynamic properties
raise to the typical properties of the local hot phase. The gravitational
softening scale within the collapsed halo is 155 pc. The final simulated objects
have structures and stellar populations consistent with observed dwarf galaxies

On one hand, as seen from Figs. \ref{scale-rel}, \ref{SSFR} and \ref{fsMh}, in  
most cases the properties of simulated (central) low-mass galaxies for a 
given \ms, in particular SSFR and \fs, agree among different authors including 
our results, in spite of the different codes used, simulations settings, and sub-grid 
physics, as well as the diversity in halo MAHs. 
On the other hand, all these simulated galaxies tend to have realistic structural,
dynamical, and gaseous properties, excepting 
the {\it too high SSFR and too low \fs\ values at a given \ms} as compared
with observational inferences out to $z\sim 1$.

As said in the Introduction, by increasing the resolution and including adequate
SF/feedback prescriptions, several of the difficulties found in previous
numerical simulations of disk galaxies have been overcome, specially the one
related to the angular momentum catastrophe. According to simulation results,
a key physical ingredient in the evolution of disk galaxies is the SF-driven 
feedback (see discussions in C10 and in the above mentioned references): 
it self-regulates SF promoting a multi-phase ISM; it removes low angular 
momentum gas from the galaxy avoiding this way the formation of a
too compact stellar component with very old stellar populations; it
drives galaxy outflows that lower the baryonic and stellar mass
fractions in the simulated galaxies. Most authors agree that
without a highly efficient SF-feedback, realistic galaxies are not reproduced. 
In our simulations, such an efficiency is attained by artificially
turning off the cooling by a time $\toff$ (only) in the cells where stars form 
 (\S\S 2.1). The times $\toff$ used here (10-40 Myr) correspond roughly 
 to the time a pressure-driven super-shell takes to reach $\gtrsim 100-150$ pc.
This latter scale corresponds roughly to our spatial resolution limits. Tests show
that for $\toff$ varying in the 10-40 Myr range, our results do not change significantly;
compare, for example, runs $E_{h1}$ and $E_{h2}$.

Regarding SF, the density and efficiency parameters used in the
SF schemes, showed to be dependent on the resolution of the
simulation \citep[e.g.,][]{Saitoh08,Brooks10} and interconnected
with the feedback and gas infall processes.  From physical
considerations (see C10), for the spatial resolutions achieved here, the 
value for $\nsf$ we adopted is $6 \ \pcc$. If $\nsf$ is increased to  $50-100 \ \pcc$
as in \citet{Governato10}, then our low-mass galaxies become too
concentrated and having too early SF (see also C10). 
 
The lack of resolution at the scales where SF and the SF-driven momentum 
and energy injection to the ISM happen, remains as a shortcoming of 
numerical simulations, in spite of the mentioned improvements in the implementation 
of sub-grid schemes. The sub-grid schemes are typically "optimized" to reproduce 
present-day galaxy properties and for  following the highest 
density regions of the simulation, where galaxies form.  Less attention
has been given to the physics of the intra-halo and intergalactic medium.
The processes related to this medium could play a relevant role
in galaxy formation and evolution, in particular at the early epochs and 
for low-mass halos.  

It is not clear whether the main physics and evolutionary
process of disk galaxy evolution are already captured by our
and other simulations and the challenge remains
only in improving resolution and tuning better the combination 
of small effects \citep[e.g.,][]{Piontek09} or some key
physical ingredients and evolutive processes are yet
missed. The answer to this question is beyond the scope
of this paper. However, the results and discussion 
presented here point out to a potential serious shortcoming
of simulations and models of low-mass central galaxies
in the context of the \lcdm\ scenario. The solution to such
a problem requires that the stellar mass assembly of 
galaxies in distinct low-mass halos be significantly
delayed to late epochs, and the smaller the galaxy, the
more should be such a delay (downsizing).

\subsection{Observational caveats}

In the last years we have seen an explosion of works aimed to determine
the stellar mass and SFR of galaxies at different redshifts (see Introduction
and Appendix for references). Yet these inferences should be taken 
with caution, though the problem stated in this paper seems to be 
robust.  Following, we discuss the main concerns about these inferences.

\subsubsection{Uncertainties in the inferences of \ms\ and SFR}

The SPS models used to fit observational data and infer hence \ms\ of galaxies
are flawed mainly by uncertainties in stellar evolution (for example in the thermally--pulsating 
asymptotic giant branch, TP-AGB, and horizontal branch phases) and by the poor 
knowledge of the initial mass function, IMF, as well as due to degeneracies like the 
one between age and metallicity (see for recent extensive discussions 
Maraston et al. 2006; Bruzual 2007; Tonini et al. 2009; Conroy, Gunn \& White 2009; 
Santini et al. 2009; Salimbeni et al. 2009). For example, 
Conroy et al. (2009) estimated that including uncertainties in stellar evolution, \ms\  at $z\sim 0$
carry errors of $\sim 0.3$ dex at 95\% confidence level with little dependence on luminosity
or color.

Maraston et al. (2006) showed that the stellar masses of galaxies with dominating stellar 
populations of $\sim 1$ Gyr age could be on average $60\%$ lower if their 
assumptions for convective overshooting during the TP--AGB phases are used.  
More recently, \citet{Salimbeni09} have determined the stellar masses of galaxies 
from a GOODS-MUSIC sample at redshifts $0.4<z<4.0$ by using
the Bruzual \& Charlot (2003, BC03) and Charlot \& Bruzual 2007 (CB07, see 
Bruzual 2007) SPS models, and the \citet{Maraston05} models (M05). 
\citet{Salimbeni09} have found that the masses obtained with MR05 and CB07 
(which take into account the TP--AGB
phase) are on average lower than those obtained with BC03. In the redshift bin 
$0.4<z<0.8$ the ratio they have found between BC03 and BC07 masses is 
$\sim 0.3$ dex for \ms$<10^{9.5}$ \msun\ and $\sim 0.05$ dex for \ms$>10^{11}$ 
\msun, while between BC03 and M05 the ratio is $\sim 0.2$ dex for all masses. 

 All the SSFR-\ms\ relations compiled here and showed in Fig. \ref{SSFR} were inferred
using the old BC03 SPS models. If we assume a conservative correction due to
the TP--AGB phase of 0.2 dex in \ms\ for all the masses and a slope of $-0.5$ 
in the logSSFR-log\ms\ relation (in most cases this slope is shallower), 
then such a relation shifts on average by 0.1 dex ($\approx 25\%$) towards 
the high-SSFR side. Therefore, if any, {\it the differences between observed and 
predicted SSFR--\ms\ relations could be even larger than showed in Fig. \ref{SSFR}}.

Regarding the determination of SFR, it could significantly depend on the used
indicator. For example, in \citet{Gilbank11} is showed that the SFRs inferred
for a sample of galaxies at $z\sim 1$ by using as a indicator the [OII] empirically
corrected for extinction as a function of \ms\ \citep{Gilbank10} are on average 
2.2 times lower than those inferred by using [OII] + 24$\mu$m, where the 
[OII] has not been corrected for extinction (since the light from SF being
reprocessed by dust should now be measured by the 24$\mu$m flux). 
This explains the difference seen in Fig. \ref{SSFR}, third panel,  
in between \citet{Gilbank11} and \citet{Noeske07} (and probably the other authors).  

Different indicators have different sensitivities to unobscured/obscured SF. 
Because observations are required at different $z$ and for different 
mass and galaxy type ranges, different indicators in the same sample are often used.
This unavoidably introduces systematics that difficult the interpretation of
correlations and evolutive trends.  
It should be said that in several of the works compiled here, the combination 
of two or more indicators that trace both obscured/unobscured SFR were 
used with calibrations given by SPS models. These 
calibrations seem to be robust to variations in the stellar evolution and extinction
assumptions, but they vary among different assumption on IMF by $\approx 
0.1-0.4$ dex. The latter systematic variations are partially compensated when calculating
the SSFR because both the SFR calibrators and the estimated \ms\  change in 
the same direction with varying the IMF. Besides, the effects of extinction
are actually minimal at low masses.

In the compilation of observational determinations showed in Fig. \ref{SSFR},  
a large diversity of SFR indicators, extinction corrections, and galaxy samples were 
used. Although there are differences in the results among these works, our 
conclusions do not change when using one or another particular work. 
This means that among all the current observational determinations, there 
is a rough convergence regarding the SSFR--\ms\ relation and its evolution.

\subsubsection{Selection effects}

Other significant sources of uncertainty and possible systematics in the inferred 
SSFR--\ms\ relations at different $z$ are (e.g., Daddi et al. 2007; Chen et al. 
2009; Stringer et al. 2010): selection effects due to sample 
incompleteness in a given wavelength; environmental effects; limit detections 
of the indicators of SFR due to flux--limits or low 
emission-line signal-to-noise ratios; in addition, obscured AGN emission could 
contaminate the infrared flux and some of the optical lines used to estimate SFR, 
though at low masses AGNs it is not frequent.  
In most of the works presented here, the authors sought to refine their samples to 
minimize the above selection effects and the contamination by AGNs.

The main concern among the mentioned above issues is that selection and 
environmental effects could bias the observed SSFR--\ms\ relations. For example, 
the high SSFRs of low--mass galaxies could be due to transient star bursts if 
the SF regime of these galaxies is dominated by episodic
processes. If among the low--mass galaxies those with low SFRs are missed due 
to detection limits, then the SSFRs of low--mass galaxies will be biased on average toward
higher SSFRs, a bias that increases for samples at higher redshifts.
Nevertheless, in most of the observational studies reporting the SSFR-\ms\ relation,
the authors discuss that while this is possible at some level, it would hardly 
change the tight SSFR--\ms\ relations observed since $z\sim 1$ 
\citep[see e.g.,][]{Noeske07a}.

On the other hand, the analysis of very local surveys helps to disentangle 
whether episodic star bursting events dominate or not the SF history of low--mass 
galaxies\footnote{Have in mind that blue, late-type galaxies are the dominant population
in the mass range $10^9 \lesssim \ms/\msun \lesssim 5\ 10^{10}$ \msun\ \citep[e.g.][]{Yang09}.}.  
From a study of the SF activity of galaxies within the 11 Mpc Local Volume,
Lee et al. (2007; see also Bothwell, Kennicutt \& Lee 2009) have found that 
intermediate--luminosity disc galaxies 
($-19 \lesssim M_B \lesssim -15$ or  $50\lesssim \vmax/\kms \lesssim 120$) 
show relatively low scatter in their SF activity, implying factors not larger than 2--3
fluctuations in their SFRs; above $\vmax\approx 120$ km/s the sequence turns off toward 
lower levels of SSFRs and larger bulge--to--disc ratios. These results are for nearby 
galaxies, where selection effects are minimal, and imply that the SSFRs of disc
galaxies with $\ms\gtrsim 5\ 10^{8}$ \msun\ follow a relatively tight sequence, without
strong fluctuations.  
For galaxies smaller than $\vmax\sim 50$ km/s (dwarfs) the situation seems different.
The results by Lee et al. (2007) show that a significant fraction of such galaxies are 
undergoing strong episodic SF fluctuations due to the large scatter in their SSFRs. 

Another observational study of nearby galaxies by James et al. (2008), also concluded that 
there is little evidence in their sample of predominantly isolated field galaxies of significant
SF through brief but intense star-burst phases. Therefore, it seems that the tight sequence 
found for normal star--forming galaxies in the SSFR--\ms\ plane in large surveys as SDSS 
(Brinchman et al. 2004; Salim et al. 2007; Schiminovich et al. 2007) is intrinsic and
due to a high degree of temporal self--regulated SF within individual galaxies. This 
sequence \citep[called the 'main sequence' in][]{Noeske07a} seems to persist back to 
$z\sim 1$ as discussed above. 

Finally, possible selection biases in the SSFR--\ms\ relation at low masses due to 
environment seem not to be a concern. For local SDSS and high $z$ (up to $z\sim 1$) 
zCOSMOS galaxy samples it was found that the relationship between SSFR and \ms\ is 
nearly the same in the highest and lowest density quartiles of {\it star-forming} 
galaxies \citep{Peng10}. What is strongly dependent on environment are the 
fractions of star-forming and quenched galaxies: the higher the environmental 
density, the higher the fraction of quenched (red) galaxies, even at lower mass.
On the average --where most galaxies live-- and low density environments, 
the fraction of red galaxies becomes smaller and smaller as the masses are lower  
\citep{Peng10}. Similar results were found by \citet{McGee10}, who determined
stellar masses and SFRs for large samples of field and group galaxies at
$z\sim 0$ by using SDDS and at $z\sim 0.4$ by using the Galaxy Environment 
Evolution Collaboration survey. The SSFR--\ms\ relation of star-forming galaxies
is consistent within the errors in the field and group environment at fixed $z$,
but the fraction of passive (red) galaxies is larger in groups than in the field
at almost all masses.
 
 \subsubsection{Concluding remarks}
 
We conclude that, in spite of significant uncertainties and possible systematics
due to selection effects, current observational determinations of the
SSFR--\ms\ relation for field galaxies at redshifts up to $z\sim 1$ have
achieved a rough consistency among them. The global empirical picture of 
the \ms\ assembly that they suggest, for sub--\mstar\ central galaxies, begins 
to be established; however, several important details such as, the separation 
by galaxy types and environment, are still highly uncertain.
According to those works, where the samples were separated into 
blue/star-forming, red/quenched  and/or AGN-dominatd galaxies 
\citep[e.g.,][]{Salim07,Bell07,Noeske07a,Karim10},  the low-mass side of 
the SSFR--\ms\ relation is actually dominated by blue/star-forming galaxies, 
which correspond typically to disk-dominated galaxies. 

Big efforts should be done in the next years to determine with better
accuracy and completeness the \ms\ and SFR of galaxies down to small masses,
up to high redshifts, and separated by galaxy type and environment. 
This implies not only observational efforts but also theoretical ones:
SPS models, used to infer the physical quantities from the
flux and spectroscopy measurements, need to be improved. The information provided by 
the SSFR as a function of \ms,
at different epochs, offers important clues for understanding how galaxies assembled
their masses and helps to constrain any theoretical approach to galaxy formation and evolution.

%=====================================================
\section{Conclusions}
%=====================================================

The N-body + Hydrodynamic ART code has been used to simulate
sub-\mstar\ central galaxies in the halo mass range of \mh($z=0$)$\sim 2-40\times 10^{10}$
\msun, varying the resolution and several of the sub-grid parameters.
Most of the obtained galaxies in our highest resolution simulations have
dynamical and structural properties in reasonable agreement with local
sub-\mstar\ field galaxies, which typically are disk, late-type galaxies
(Figs. \ref{scale-rel} and \ref{Vevol}, and Table \ref{prop}). 

Although we have simulated only 10 different galaxies, all immersed in
different {\it distinct} halos, 
our results allows us to answer, at least preliminary, the main questions 
stated in the Introduction and hence conclude that:

 \begin{itemize}
 
\item The SSFRs at  $z\sim 0$ and $\sim 0.3$ of simulated galaxies in the mass range 
$\ms\sim 10^8-10^{10}$  \msun\ are $5-10$ times lower than the mean values inferred 
from several observational samples of field galaxies (Fig. \ref{SSFR}). 
There is not a high-resolution simulation with a SSFR value high enough  
to lie above the lower $1\sigma$ scatter of the observational determinations by
\citet{Salim07} of star-forming galaxies at $z\sim 0$ (most observed low-mass central galaxies are 
actually star-forming).  At higher redshifts, $z\sim 1.0-1.5$, most simulated galaxies 
have masses below the completeness limits of current observational inferences at those 
$z$. However, both the measurements of the SSFRs of those galaxies 
which are below these limits and the extrapolations to lower masses of the 
SSFR--\ms\ relations of the complete samples, are higher than the SSFRs
of our simulated galaxies, though the differences are apparently smaller than at 
$z\sim 0$.  Several of the simulations at $z\sim 1.0-1.5$ have already 
SSFRs around the value corresponding to a galaxy forming stars with a rate equal 
to its past average. On  the contrary, at $z\lesssim 0.3$, most simulations have 
SFR values well below this case, in clear disagreement with most current observational 
determinations.  

\item The evolution of the observationally determined SSFR--\ms\ relations
shows a clear trend of downsizing for the typical stellar mass, $M_{\rm quench}$,
of galaxies that transit from active to passive (those that cross below the line of 
constant SFR at a given $z$ in Fig. \ref{SSFR}).  There is not any evidence of 
such a downsizing trend in the simulations. 

\item The stellar mass fractions ($\fs\equiv \ms/\mh$) of simulated galaxies are $5-10$ times
larger at $z\sim 0$ than current  determinations (semi-empirical and direct)
of these fractions
as a function of \ms\  (Fig. \ref{fsMh}).  At higher redshifts, the differences, at a given mass, 
increase even more; at $z=1.5$, the \fs\ of simulated galaxies are around 30 times larger 
than the semi-empirical inferences or their extrapolations to lower masses. Put in another way, 
while the \fs--\mh\ relation of simulated galaxies would tend to remain constant or slightly shift to 
higher \fs\ values as $z$ increases, the current semi-empirical inferences show that
the low-mass side of this relation shifts significantly to lower \fs\ values (see also 
Fig. \ref{fsMh_evol}).  

\end{itemize}

Therefore, our numerical simulations confirm the issues previously found with 
semi-numerical models of disk galaxy evolution (and eventually SAMs) in the 
context of the \lcdm\ scenario: {\it low-mass disk-like galaxy models seem to assemble their 
stellar masses on average much earlier than suggested by several pieces of evidence
such as the observationally inferred SSFR--\ms\ and \fs--\ms\ relations at redshifts $z<1$. }
As pointed out in \citet{FAR10} and FA10, the \ms\ assembly of low-mass
modeled galaxies follow, in first instance, the mass assembly of their
corresponding halos, and in the \lcdm\ scenario, low-mass halos assemble
on average earlier than more massive halos (upsizing). The different astrophysical 
processes followed in current models and numerical simulations
of galaxy formation and evolution (gas cooling and infall, SF, SF feedback,
SN-driven galaxy outflows, etc.) makes the growth of \ms\ deviates from 
that of the corresponding dark halo but not as strongly as observations 
apparently suggest (downsizing; see  Fig. 4 in FA10 for a comparison of the average 
trends of halo MAHs and empirically inferred average galaxy \ms\ tracks).

If the empirical picture of sub-$\mstar$ central galaxies assembly is confirmed
(see \S\S 4.2 for a discussion in current uncertainties), 
then the \ms\ growth of central galaxies less massive
than \ms($z=0$)$\sim 3\times 10^{10}$ \msun\ seems to require a significant
delay with respect to the evolution of their corresponding \lcdm\ halos;
besides, the smaller the galaxy the longer such a delay 
\citep[see also][]{Noeske07}.  
According to \citet{Bouche+10}, this
mass-dependent delay in the \ms\ assembly can be explained by a halo mass floor 
 $M_{\rm min}\approx 10^{11}$ \msun, below which the halo-driven gas accretion
 is quenched.

\section*{Acknowledgments}

We thank the Referee for his/her comments and suggestions that improved
the presentation of the paper.
We are grateful to A. Kravtsov for providing us with the numerical code and to N. Gnedin 
for providing us the analysis and graphics package IFRIT.
The authors acknowledge PAPIIT-UNAM grants IN114509 to V.A. and IN112806 to P.C.
and CONACyT grant 60354 (to V.A.,  P.C. and O.V).
Some of the simulations presented in this paper were performed on the HP CP 
4000 cluster (Kan-Balam) at DGSCA-UNAM.

%=============================

\appendix

\section{The compilation of observational works}

In Fig. \ref{SSFR} the simulation results in the SSFR--\ms\ plane are compared with a large
body of observational inferences at four $z$ bins compiled from the 
literature.  Following, a brief description of these inferences at each $z$ bin is presented.

{\it a) $z=0$ panel:} The linear fit to the SSFR--\ms\ relation and its  
intrinsic width as determined in \citet{Salim07} for the sub-sample 
of normal star-forming galaxies (optical emission lines are used for
classifying galaxies) in a volume-corrected sample of around 50,000 
SDSS galaxies are shown. The dust-corrected SFRs were obtained 
by fitting ({\it GALEX}) UV and SDSS photometry to a library of 
dust-attenuated SPS models. While the fraction of galaxies with 
negligible H$\alpha$ emission-line detection in their sample is significant, 
most of them are luminous (massive) ones. Therefore, their SSFR--\ms\ 
relation for 'star-forming galaxies' at low masses ($\ms\lesssim 2\times 
10^{10}$ \msun) is almost the same if galaxies with no H$\alpha$ 
detection are included (compare the mentioned fit with the one 
carried out by the same authors for the {\it entire} sample, where a 
Schechter function was used for the fitting,  long-dashed line; the 
characteristic mass is $M_0=10^{11.03}$ \msun). Results similar to 
\citet{Salim07} were found by other authors \citep[e.g.,][]{Schiminovich07}.

{\it b) $z=0.33$ and $z=1.00$ panels:} The median and standard deviation of 
$\log$SSFR vs $\log$\ms\ of the sequence of star-forming field galaxies as reported 
in \citet{Noeske07} are plotted in these panels (crosses with vertical error bars).  
The data used by these authors consisted of 2905 galaxies out to $z=1.1$ from 
the All-Wavelength Extended Groth Strip International Survey (AEGIS). The SFRs
for galaxies with robust 24 $\mu$m detections were derived from 24 $\mu$m 
luminosity + non-extinction corrected emission lines, and for galaxies below
the 24 $\mu$m detection, from extinction-corrected emission lines.
The symbols show only the \ms\ range where the sample is $>95\%$ complete 
in the redshift bins $0.20\le z<0.45$ and $0.85\le z<1.10$, respectively. For 
masses smaller than the sample completeness ($\ms\sim 10^8$ \msun\ and 
$\sim 10^9$ \msun, respectively) the derived SSFRs continue to increase 
on average for lower values of \ms. Around 30\% of galaxies in their sample at 
all $z$ have not robust 24 $\mu$m or emission-line detections, which implies
very low SFRs. The inclusion of these galaxies in the SSFR-\ms\ relation,
decreases the overall median SSFRs and it would be lower than the one
shown in Fig. \ref{SSFR}.
However, almost all of these galaxies ($>90\%$) are in the red sequence and
have early-type morphologies, being besides relatively massive \citep{Noeske07a,Noeske07}. 

In these panels are also plotted the 
average data (from stacking) taken from the SSFR vs \ms\ plots at $0.2<z\le 0.4$ 
and $0.8<z\le 1.0$ given in \citet{Bell07} for 'all galaxies' (open stars) 
and for blue galaxies only (blue asterisks). Results 
correspond to the Chandra Deep Field South (CDFS) sample; the COMBO-17 
survey in conjunction with \textit{Spitzer} 24$\mu$m data were used for
estimating \ms\ and SFR (for the latter, UV and 24$\mu$m luminosities
were used). In the 
low-mass side ($\ms< 3\ 10^{10}$ \ms) and for all $z$ bins,
blue galaxies dominate in such a way that the average SSFRs for 
the total sample (blue + red galaxies) is close to the average 
of the blue cloud sample. The mass completeness limits reported
in \citet{Bell07} are indicated with the dotted vertical arrows.

{\it c) $z=0.33$, $z=1.00$, and $z=1.50$ panels:} \citet{Santini09} presented 
least-square linear fits to the SFR--\ms\ relations (SFR=A$\ms^\beta$) obtained
from the GOOD-MUSIC catalog, which has multi-wavelength coverage from 0.3 
to 24 $\mu$m (complete SED fittings and UV--24 $\mu$m luminosities were
used for estimating SFRs). We plot in Fig. \ref{SSFR} these fits but divided by \ms\ (blue lines)
for the redshift intervals $0.3\le z< 0.6$ (second panel), $0.6\le z<1.0$ and 
$1.0\le z<1.5$ (third panel), and $1.0\le z<1.5$ and $1.5\le z<2.5$ (fourth panel). 

Similarly, the fits to the SSFR-\ms\ relations obtained in \citet{Rodighiero10} 
for the redshift bins $0.3\le z< 0.6$ (second panel), $0.6\le z< 1.0$ and $1.0\le z< 1.5$
(third panel), and $1.0\le z< 1.5$ and $1.5\le z< 2.5$ (fourth panel) 
are plotted (black solid lines). These authors used deep observations of the 
GOODS-N field taken with PACS on board of the \textit{Herschel} satellite,
which allows them to robustly derive the total infrared luminosity of galaxies,
and combine with the ancillary UV luminosities in order to calculate the SFRs. 
In both cases \citet{Santini09} and \citet{Rodighiero10}, we plot their
fits only for masses where samples are complete.

 %%%%%
The linear fits to radio-{\it stacking}-based measurements of the SSFR as 
a function of \ms\ recently presented by \citet{Karim10} are also plotted
in these three panels (long-dashed red line).  
The fits correspond to star-forming galaxies from a deep 3.6--$\mu$m--selected
sample of $> 10^5$ galaxies ($0.2<z<3$) in the 2 deg$^2$ COSMOS field, 
and are shown only for galaxies above the completeness
limit at the given $z$.  The fits to the entire sample are steeper than those to only
star-forming galaxies, in such a way that at low masses both intersect, but at larger
masses, the former lie below the latter. 
The image-stacking technique applied in order to increase the signal-to-noise ratio 
in the VLA 1.4 GHz radio continuum observations allows one to estimate only average SFRs
of the stacked population; they cannot shed light on the intrinsic dispersion of 
individual sources.
%%%%%

Also shown in these panels are the average SSFRs corresponding to the
lowest mass bins --where given samples are still complete-- from: 

{\it Damen et al. (2009b).-} The mass bins of 0.5 dex
width in each panel correspond to $z\approx 0.4, 1.0$ and 1.4, respectively;
the horizontal error bar indicates the width of the \ms\ bin, 0.5 dex, and the 
vertical error bar represents the bootstrapped 68\% confidence level on the 
average SSFR in the bin; the FIREWORKS catalog for the GOODS-CDFS 
generated by \citet{Wuyts08} was used to infer \ms\ and SFRs (infrared 
and UV luminosities were combined to derive the latter and no selection on 
color/morphology was applied; i.e., all galaxies were included). 

{\it Dunne et al. (2009).-} The lowest complete mass bins of 1 dex width correspond to  $z=0.50$, 
0.95, and 1.40, respectively, with the average SSFRs and their scatters 
shown with the error bars centered in the median value of the mass bin;
stacked deep radio mosaics of $K-$band selected galaxies from the 
UKIDSS were used to determine \ms\ and SFRs --rest-frame 1400-MHz
luminosities were used to derive the latter; the case using the \citet{Bell03}
conversion and applied to the whole sample is reproduced here.

{\it Kajisawa et al. (2010).-} The lowest complete mass bins of 0.5 dex width
correspond to the median dust-corrected SFRs, where the rest-frame 2800 $\AA$ 
luminosity was used as indicator. This UV luminosity was estimated by the best-SED 
fitting of the multi-band photometry ({\it UBVizJHK}, 3.6 $\mu$m, 4.5 $\mu$m, and 5.8 $\mu$m) 
with the BC03 models. We reproduce the reported values at redshift bins 
of $0.5<z<1.0$, and $1.0<z<1.5$.

\begin{table*}
\begin{center}
 \caption{Summary of sources and corrections}
 \begin{tabular}{@{}lcccc@{}}\\
 \hline
  Source  & SFR tracer & IMF   & \ms\ offset\tablenote{Shift in dex to convert \ms\ to the Chabrier IMF }  & 
  SFR offset\tablenote{Shift in dex to convert SFR to the Chabrier IMF}  \\
 \hline
Salim et al. (2007)     & $L_{UV}$, $H_{\alpha}$ & Chabrier (2003) & 0.0  & 0.0 \\
Bell et al. (2007)      & $L_{IR} + L_{UV}$    & Chabrier (2003) & 0.0  & 0.0 \\
Noeske et al. (2007)    & $L_{24\mu m} +$ emission lines & Kroupa (2001) & 0.0
 & 0.0 \\
Santini et al. (2009)   & $L_{IR} + L_{UV}$    & Salpeter (1955) & -0.25 & -0.176 \\
Damen et al. (2009)     & $L_{IR} + L_{UV}$    & Kroupa (2001)   & 0.0   & 0.0 \\
Dunne et al. (2009)     & $L_{1400 MHz} $      & Salpeter (1955) & -0.25 & -0.176 \\
Kajisawa et al. (2010)  & $L_{UV}$(corrected)  & Salpeter (1955) & -0.25 &
-0.176 \\
Karim et al. (2010)     &  $L_{1400 MHz}$      & Chabrier (2003) & 0.0   &  0.0  \\
Rodighiero et al. (2010)& $L_{IR} + L_{UV}$    & Salpeter (1955) & -0.25 & -0.176 \\
Gilbank et al. (2011)   & [OII]-($H_{\alpha}$)   & Baldry \& Glazebrook (2003) &
-0.08  &   -0.18 \\
\hline
\end{tabular}
\label{appendix}
\end{center}
\end{table*}

After the completion of this paper, appeared the work by \citet{Gilbank11},
where the SSFR--\ms\ relation for star-forming galaxies at $z\sim 1$ is presented 
down to masses smaller than those previously determined (their linear fit is plotted in 
panel $z=1$ of Fig. \ref{SSFR} with a long-dashed purple line).  For the low-mass 
side, they analyzed a sample of 199 low-mass galaxies ($10^{8.5}\lesssim \ms/\msun\lesssim 
10^{9.3}$) from the field CDFS, with spectroscopy redshifts determined in the 
range $0.88 < z \le 1.15$ and a sampling completeness $> 80\%$ \citep{Gilbank10}. 
Stellar masses were determined by SED-fitting of the photometry, at the 
given $z$, using a grid of PEGASE.2 models.
The [OII] line is used as  the SFR indicator (actually the Kennicutt 1998 
SFR--H$\alpha$ relation is used by assuming an [OII]/H$\alpha$ ratio of 0.5), 
empirically corrected as a function
of \ms\ in order to correct for extinction and other systematic effects. Their correction
implies on average a lower SFR than previously determined, by using for example
[OII](non-corrected) + $24\mu$m (see for a discussion \S\S 4.2.1). The relation
by \citet{Gilbank11} is indeed the one that implies the lowest estimates for SSFR among
all those compiled in Fig. \ref{SSFR} at $z\approx 1$. These authors estimated
also the local SSFR--\ms\ of blue galaxies by using SDSS Stripe 82  and the H$\alpha$
line as the SFR indicator. The linear fit to their results lies below the average relation
reported by \citet[][by $\approx 0.25$ dex at $10^9$ \msun]{Salim07}.

In Table \ref{appendix}, a summary of all the sources used in Fig. \ref{SSFR} and
described above is presented. In all the cases, $h=0.70$ or 0.71 was assumed.
Columns 2 and 3 report the main SFR indicator(s) 
and IMF used in each source, respectively. Columns 4 and 5 give the average shifts
in dex applied here respectively to \ms\ and SFR in order to correct for a \citet{Chabrier03} IMF
when necessary.  According to \citet[][]{Bell07}, the stellar masses and SFRs calculated with 
\citet{Chabrier03} and \citet{Kroupa} IMFs (when using $L_{IR} + L_{UV}$ as indicator)
are consistent to within $\lesssim 10\%$, so we do no apply corrections for those
cases when a \citet{Kroupa} IMF was used (see for similar conclusions Karim et al. 2010 
and Salim et al.  2007 for the $L_{1400 MHz}$ and $H_{\alpha}$ indicators , respectively). 
For the corrections in cases when the \citep{Salpeter} IMF was used, see \citet{Santini09}. For 
the \citet{BG03} IMF, we have used the corrections given in \citet{Gilbank11} for
passing to a \citet{Kroupa} IMF.  We are aware that all 
these corrections are yet uncertain, and could actually vary from indicator
to indicator and with $z$. In any case, they are relatively small, in particular for the SSFR.


\begin{thebibliography}{}
\bibitem[Agertz et al.(2011)]{Agertz11} Agertz, O., Teyssier, R., \& Moore, B.\ 2011, \mnras, 410, 1391
\bibitem[Avila-Reese et al.(2008)]{Avila08} Avila-Reese, V., 
Zavala, J., Firmani, C., \& Hern{\'a}ndez-Toledo, H.~M.\ 2008, \aj, 136, 1340 
\bibitem[Baldry \& Glazebrook(2003)]{BG03} Baldry, I.~K., \& Glazebrook, K.\ 2003, \apj, 593, 258 
\bibitem[Baldry et al.(2004)]{B04} Baldry, I.~K., Glazebrook, K., Brinkmann, J., Ivezi{\'c}, {\v Z}., Lupton, R.~H., Nichol, 
R.~C., \& Szalay, A.~S.\ 2004, \apj, 600, 681 
\bibitem[Bauer et al.(2005)]{Bauer05} Bauer, A.~E., Drory, N., Hill, G.~J., \& Feulner, G.\ 2005, \apjl, 621, L89
\bibitem[Behroozi et al.(2010)]{BCW10} Behroozi, P.~S., Conroy, C., \& Wechsler, R.~H.\ 2010, \apj, 717, 379
\bibitem[Bekki(2008)]{Bekki08} Bekki, K.\ 2008, \apjl, 680, L29 
\bibitem[Bell et al.(2003)]{Bell03} Bell, E.~F., McIntosh, D.~H., Katz, N., \& Weinberg, M.~D.\ 2003, \apjl, 585, L117 
\bibitem[Bell et al. (2007)]{Bell07} Bell E.~F., Zheng X.~Z., Papovich C., Borch A., Wolf C., Meisenheimer K., 
         2007, ApJ, 663, 834 
\bibitem[Benson et al.(2003)]{Benson03} Benson, A.~J., Bower, 
R.~G., Frenk, C.~S., Lacey, C.~G., Baugh, C.~M., \& Cole, S.\ 2003, \apj, 599, 38 
\bibitem[Bertone et al.(2007)]{Bertone07} Bertone, S., De Lucia, G., \& Thomas, P.~A.\ 2007, \mnras, 379, 1143
\bibitem[Bothwell et al.(2009)]{2009MNRAS.400..154B} Bothwell, M.~S., 
Kennicutt, R.~C., \& Lee, J.~C.\ 2009, \mnras, 400, 154
\bibitem[Bouch{\'e} et al.(2010)]{Bouche+10} Bouch{\'e}, N., et al.\ 2010, \apj, 718, 1001
 \bibitem[Bower et al.(2006)]{Bower06} Bower, R.~G., Benson, 
A.~J., Malbon, R., Helly, J.~C., Frenk, C.~S., Baugh, C.~M., Cole, S., 
\& Lacey, C.~G.\ 2006, \mnras, 370, 645 
\bibitem[Brinchmann et al.(2004)]{2004MNRAS.351.1151B} Brinchmann, J., 
Charlot, S., White, S.~D.~M., Tremonti, C., Kauffmann, G., Heckman, T., 
\& Brinkmann, J.\ 2004, \mnras, 351, 1151 
\bibitem[Brooks et al.(2011)]{Brooks10} Brooks, A., et al.\  2011, \apj, 728, 51
\bibitem[Bruzual \& Charlot(2003)]{2003MNRAS.344.1000B} Bruzual, G., \& Charlot, S.\ 2003, \mnras, 344, 1000 
\bibitem[Bruzual(2007)]{2007ASPC..374..303B} Bruzual, G.\ 2007, From Stars 
to Galaxies: Building the Pieces to Build Up the Universe, in "Astronomical Society of the Pacific 
Conference Series", Eds. A.~Vallenari, R.~Tantalo, L.~Portinari, \& A.~Moretti, 374, 303 
\bibitem[Bundy et al.(2006)]{Bundy06} Bundy, K., et al.\ 2006, \apj, 651, 120 
\bibitem[Ceverino \& Klypin (2009)]{CK2009} Ceverino, D., \& Klypin, A. 2009, ApJ, 695, 292
\bibitem[Chabrier(2003)]{Chabrier03} Chabrier, G.\ 2003, \pasp, 115, 763 
\bibitem[\protect\citeauthoryear{Chen et al.}{2009}]{Chen09} Chen Y.-M., Wild V., Kauffmann G., Blaizot J., 
Davis M., Noeske K., Wang J.-M., Willmer C., 2009, MNRAS, 393, 406 
\bibitem[Col\'{\i}n et al.(2010)]{Colin10} Col\'{\i}n, P., Avila-Reese, 
V., V\'azquez-Semadeni, E., Valenzuela, O., \& Ceverino, D.\ 2010, \apj, 713, 535
\bibitem[Conroy et al.(2009)]{Conroy09a} 
         Conroy, C., Gunn, J.~E., \& White, M.\ 2009, \apj, 699, 486 
\bibitem[Conroy \& Wechsler(2009)]{CW09} Conroy, C., \& Wechsler, R.~H.\ 2009, \apj, 696, 620
\bibitem[\protect\citeauthoryear{Cowie et al.}{1996}]{Cowie96} 
Cowie L.~L., Songaila A., Hu E.~M., Cohen J.~G., 1996, AJ, 112, 839 
\bibitem[Croton et al.(2006)]{Croton06} Croton, D.~J., et al.\  2006, \mnras, 367, 864 
\bibitem[Daddi et al.(2007)]{Daddi07} Daddi, E., et al.\ 2007, \apj, 670, 173 
\bibitem[Damen et al.(2009a)]{Damen09a} 
         Damen, M., Labb{\'e}, I., Franx, M., van Dokkum, P.~G., Taylor, E.~N., 
         \& Gawiser, E.~J.\ 2009a, \apj, 690, 937 
\bibitem[Damen et al.(2009b)]{Damen09b} Damen, M., F{\"o}rster Schreiber, N.~M., Franx, M., 
         Labb{\'e}, I., Toft, S., van Dokkum, P.~G., \& Wuyts, S.\ 2009b, \apj, 705, 617 
\bibitem[De Lucia et al.(2004)]{deLucia04} De Lucia, G., Kauffmann, G., \& White, S.~D.~M.\ 2004, \mnras, 349, 1101 
\bibitem[De Lucia et al.(2006)]{DeLucia06} De Lucia, G., Springel, V., White, S.~D.~M., Croton, D., 
\& Kauffmann, G.\ 2006, \mnras, 366, 499
\bibitem[De Rijcke et al.(2007)]{deRijcke07} De Rijcke, S., Zeilinger, W.~W., Hau, G.~K.~T., Prugniel, P., 
\& Dejonghe, H.\ 2007, \apj, 659, 1172 
\bibitem[de Rossi et al.(2010)]{derossi10} de Rossi, M.~E., Tissera, P.~B., \& Pedrosa, S.~E.\ 2010, \aap, 519, A89
\bibitem[Drory \& Alvarez(2008)]{DroryAlvarez} Drory, N., \& Alvarez, M.\ 2008, \apj, 680, 41
\bibitem[Drory et al.(2009)]{Drory+2009} Drory, N., et al.\ 2009, \apj, 707, 1595 
\bibitem[Dunne et al.(2009)]{2009MNRAS.394....3D} Dunne, L., et al.\ 2009, \mnras, 394, 3 
\bibitem[Dutton \& van den Bosch(2009)]{Dutton09} Dutton, A.~A., \& van den Bosch, F.~C.\ 2009, \mnras, 396, 141
\bibitem[Dutton et al.(2010a)]{Dutton10a} Dutton, A.~A., van den Bosch, F.~C., \& Dekel, A.\ 2010a, \mnras, 405, 1690 
\bibitem[Dutton et al.(2010b)]{Dutton10c} Dutton, A.~A., et al.\  2010b, \mnras, 1493
\bibitem[Elbaz et al.(2007)]{Elbaz07} Elbaz, D., et al.\ 2007, \aap, 468, 33
\bibitem[Fakhouri et al.(2010)]{2010MNRAS.406.2267F} Fakhouri, O., Ma, 
C.-P., \& Boylan-Kolchin, M.\ 2010, \mnras, 406, 2267 
\bibitem[Ferland et al. (1998)]{Ferland98} Ferland, G.J., Korista, K.T., Verner, D.A., Ferguson, J.W., Kingdon, J.B.,  \& Verner, E.M. 1998, PASP, 110, 761
\bibitem[Feulner et al.(2005)]{Feulner+05} Feulner, G., Gabasch, 
A., Salvato, M., Drory, N., Hopp, U., \& Bender, R.\ 2005, \apjl, 633, L9
\bibitem[Firmani \& Avila-Reese(2000)]{FA00} Firmani, C., \& Avila-Reese, V.\ 2000, \mnras, 315, 457
\bibitem[Firmani \& Avila-Reese(2010)]{FA10} Firmani, C., \& Avila-Reese, V.\ 2010, \apj,  723, 755  (FA10)
\bibitem[Firmani et al.(2010)]{FAR10} Firmani, C., Avila-Reese, V., \& Rodr\'iguez-Puebla,
A. \ 2010, \mnras, 404, 1100 
\bibitem[Fontanot et al.(2009)]{Fontanot09} Fontanot, F., De Lucia, G., Monaco, P., Somerville, R.~S., \& Santini, P.\ 
 2009, preprint (astro-ph/0901.1130)
\bibitem[Geha et al.(2006)]{Geha06} Geha, M., Blanton, M.~R., Masjedi, M., \& West, A.~A.\ 2006, \apj, 653, 240 
\bibitem[Gibson et al.(2009)]{Gibson09} Gibson, B.~K., Courty, S., S{\'a}nchez-Bl{\'a}zquez, P., Teyssier, R., House, E.~L., Brook, C.~B.,  \& Kawata, D.\ 2009, IAU Symposium, 254, 445 
\bibitem[Gilbank et al.(2011)]{Gilbank11} Gilbank, D.~G., et al.\  2011, \mnras, 402, in press, arXiv:1101.3780
\bibitem[Gilbank et al.(2010)]{Gilbank10} Gilbank, D.~G., et al.\ 2010, \mnras, 405, 2419 
\bibitem[Governato et al.(2007)]{Governato07}
	 Governato, F., Willman, B., Mayer, L., Brooks, A., Stinson, G., Valenzuela, O.,
         Wadsley, J., \& Quinn, T. 2007, \mnras, 374, 1479
\bibitem[Governato et al.(2010)]{Governato10} Governato, F., et al.\ 2010, \nat, 463, 203  
\bibitem[Guo 
\& White(2008)]{2008MNRAS.384....2G} Guo, Q., \& White, S.~D.~M.\ 2008, \mnras, 384, 2 
\bibitem[Guo et al.(2010)]{Guo10} Guo, Q., White, S., Li, C., \& Boylan-Kolchin, M.\ 2010, \mnras, 404, 1111         
\bibitem[Haardt \& Madau (1996)]{HM96}  Haardt, F., \& Madau, P. 1996, \apj, 461, 20
\bibitem[Hopkins et al.(2007)]{Hopkins07} Hopkins, P.~F., Bundy, K., Hernquist, L., \& Ellis, R.~S.\ 2007, \apj, 659, 976
\bibitem[James et al.(2008)]{2008A&A...484..703J} James, P.~A., Prescott, M., \& Baldry, I.~K.\ 2008, \aap, 484, 703 
\bibitem[Kajisawa et al.(2009)]{Kajisawa09} Kajisawa, M., et al.\  2009, \apj, 702, 1393
\bibitem[Kajisawa et al.(2010)]{Kajisawa10} Kajisawa, M., Ichikawa, T., Yamada, T., Uchimoto, Y.~K., Yoshikawa, 
T., Akiyama, M., \& Onodera, M.\ 2010, \apj, 723, 129 
\bibitem[Karim et al.(2011)]{Karim10} Karim, A., et al.\ 2011, \apj, 730, 61
\bibitem[Kennicutt(1998)]{1998ARA&A..36..189K} Kennicutt, R.~C., Jr.\ 1998, \araa, 36, 189 
\bibitem[Kere{\v s} et al.(2009)]{Keres09} Kere{\v s}, D., Katz, N., Dav{\'e}, R., Fardal, M., 
\& Weinberg, D.~H.\ 2009, \mnras, 396, 2332 
\bibitem[Klypin \& Holtzman (1997)]{kh97}  Klypin, A., \& Holtzman, J. 1997, preprint (astro-ph/9712217)
\bibitem[Klypin et al. (2001)]{KKBP01} Klypin, A., Kravtsov, A.V.,
         Bullock, J.S., \& Primack, J.R. 2001, ApJ, 554, 903 
\bibitem[Klypin et al. (2010)]{Klypin10} Klypin, A., Trujillo-Gomez, S., \& Primack, J. 2010, arXiv:1002.3660
\bibitem[Kravtsov et al.(1997)]{KKK97}    Kravtsov, A.V., Klypin, A.A., \& Khokhlov, A.M., 1997,  \apjs, 111, 73
\bibitem[Kravtsov(2003)]{Kravtsov03}  Kravtsov, A.V. 2003, \apjl, 590, L1
\bibitem[Kravtsov et al.(2005)]{KNV05} Kravtsov, A.V., Nagai, D., \& Vikhlinin, A.A. 2005, \apj, 625, 588
\bibitem[Kroupa(2001)]{Kroupa} Kroupa, P.\ 2001, \mnras, 322, 231 
\bibitem[Lee et al.(2007)]{2007ApJ...671L.113L} Lee, J.~C., Kennicutt, R.~C., Funes, S.~J., Jos{\'e} G., Sakai, S., 
\& Akiyama, S.\ 2007, \apjl, 671, L113
\bibitem[Li et al.(2008)]{2008MNRAS.389.1419L} Li, Y., Mo, H.~J., 
\& Gao, L.\ 2008, \mnras, 389, 1419 
\bibitem[Liu et al.(2010)]{Liu10} Liu, L., Yang, X., Mo, H.~J., van den Bosch, F.~C., \& Springel, V.\ 
2010, \apj, 712, 734
\bibitem[Maraston(2005)]{Maraston05} Maraston, C.\ 2005, \mnras, 362, 799 
\bibitem[\protect\citeauthoryear{Maraston et al.}{2006}]{2006ApJ...652...85M} Maraston C., Daddi E., Renzini A., Cimatti 
A., Dickinson M., Papovich C., Pasquali A., Pirzkal N., 2006, ApJ, 652, 85 
\bibitem[Martig et al.(2009)]{Martig09} Martig, M., Bournaud, F., Teyssier, R., \& Dekel, A.\ 2009, \apj, 707, 250 
\bibitem[Mayer et al.(2008)]{Mayer08} Mayer, L., Governato, F., \& Kaufmann, T.\ 2008, Advanced Science Letters, 1, 7 
\bibitem[Miller \& Scalo (1979)]{MS79}  Miller, G.E., \& Scalo, J.M. 1979, \apjs, 41, 513
\bibitem[McGaugh(2005)]{Mcgaugh05} McGaugh, S.~S.\ 2005, \apj, 632, 859
\bibitem[McGee et al.(2010)]{McGee10} McGee, S.~L., Balogh, M.~L., Wilman, D.~J., 
Bower, R.~G., Mulchaey, J.~S., Parker, L.~C., \& Oemler, A., Jr.\ 2010, \mnras, in press, arXiv:1012.2388 
\bibitem[More et al.(2010)]{More10} More, S., van den Bosch, 
F.~C., Cacciato, M., Skibba, R., Mo, H.~J., \& Yang, X.\ 2010, \mnras, 1464 
\bibitem[Mortlock et al.(2011)]{Mortlock11} Mortlock, A., Conselice, C.~J., Bluck, A.~F.~L., Bauer, A.~E., Gruetzbauch, R., Buitrago, F., \& Ownsworth, J.\ 2011, \mnras, in press, arXiv:1101.2867 
\bibitem[Moster et al.(2010)]{Moster10} Moster, B.~P., Somerville, R.~S., Maulbetsch, C., van den Bosch, F.~C., 
Macci{\`o}, A.~V., Naab, T., \& Oser, L.\ 2010, \apj, 710, 903 
\bibitem[Naab et al.(2007)]{Naab07} Naab, T., Johansson, P.~H., Ostriker, J.~P., \& Efstathiou, G.\ 2007, \apj, 658, 710
\bibitem[Neistein et al.(2006)]{Neisten06} Neistein, E., van den Bosch, F.~C., \& Dekel, A.\ 2006, \mnras, 372, 933 
\bibitem[Noeske et al.(2007a)]{Noeske07a} Noeske, K.~G., et al.\  2007a, \apjl, 660, L43 
\bibitem[Noeske et al. (2007b)]{Noeske07}  Noeske K.~G., et al., 2007b, ApJ, 660, L47
\bibitem[Oliver et al.(2010)]{Oliver10} Oliver, S., et al.\ 2010, \mnras, 405, 2279
\bibitem[Oppenheimer \& Dav{\'e}(2008)]{Oppenheimer08} Oppenheimer, B.~D., \& Dav{\'e}, R.\ 2008, \mnras, 387, 577 
\bibitem[Oppenheimer et al.(2010)]{Oppenheimer10} Oppenheimer, B.~D., 
Dav{\'e}, R., Kere{\v s}, D., Fardal, M., Katz, N., Kollmeier, J.~A., \& Weinberg, D.~H.\ 2010, \mnras, 406, 2325 
\bibitem[Pasquali et al.(2010)]{Pasquali10} Pasquali, A., 
Gallazzi, A., Fontanot, F., van den Bosch, F.~C., De Lucia, G., Mo, H.~J., \& Yang, X.\ 2010, \mnras, 407, 937 
\bibitem[Peng et al.(2010)]{Peng10} Peng, Y., et al.\ 2010, \apj, 721, 193 
\bibitem[Piontek \& Steinmetz(2011)]{Piontek09} Piontek, F., \& Steinmetz, M. \ 2011, \mnras, 410, 2625
\bibitem[Pozzetti et al.(2009)]{Pozzetti10} Pozzetti, L., et al.\ 2010, \aap, 523, A13 
\bibitem[Rodighiero et al.(2010)]{Rodighiero10} Rodighiero, G., et al.\ 2010, \aap, 518, L25
\bibitem[ Rodr\'iguez-Puebla et al. (2011)]{R11} Rodr\'iguez-Puebla,A., Avila-Reese, V.,  Firmani, C., \&
Col\'{\i}n, P. \ 2010, RevMexAA, 48, in press, arXiv:1103.4151
\bibitem[Saitoh et al.(2008)]{Saitoh08} Saitoh, T.~R., Daisaka, H., Kokubo, E., 
         Makino, J., Okamoto, T., Tomisaka, K., Wada, K., \& Yoshida, N.\ 2008, PASJ, 60, 667 
\bibitem[\protect\citeauthoryear{Salim et al.}{2007}]{Salim07}  Salim S., et al., 2007, ApJS, 173, 267 
\bibitem[Salimbeni et al.(2009)]{Salimbeni09} Salimbeni, S., Fontana, A., Giallongo, E., Grazian, A., 
Menci, N., Pentericci, L., \& Santini, P.\ 2009, American Institute of Physics Conference Series, 1111, 207 
\bibitem[Salpeter(1955)]{Salpeter} Salpeter, E.~E.\ 1955, \apj, 121, 161 
\bibitem[Santini et al.(2009)]{Santini09} Santini, P., et al.\ 2009, \aap, 504, 751 
\bibitem[Sawala et al.(2011)]{Sawala10} Sawala, T., Guo, Q., 
Scannapieco, C., Jenkins, A., \& White, S.~D.~M.\ 2011, \mnras, 64  
\bibitem[Scannapieco et al.(2006)]{Scannapieco06} Scannapieco, C., Tissera, P.~B., White, S.~D.~M., 
\& Springel, V.\ 2006, \mnras, 371, 1125 
\bibitem[Scannapieco et al.(2008)]{Scannapieco08} 
        Scannapieco, C., Tissera, P.~B., White, S.~D.~M., \& Springel, V.\ 2008,  \mnras, 389, 1137
\bibitem[Schiminovich et al.(2007)]{Schiminovich07} Schiminovich, D., et al.\ 2007, \apjs, 173, 315 
\bibitem[Somerville et al.(2008)]{Somerville08}  Somerville, R.~S., et al.\ 2008, \apj, 672, 776
\bibitem[Springel et al.(2001)]{Springel01} Springel, V., White, S.~D.~M., Tormen, G., \& Kauffmann, G.\ 2001, \mnras, 328, 726 
\bibitem[Stinson et al.(2009)]{Stinson09}  Stinson, G.~S., Dalcanton, J.~J., Quinn, T., Gogarten, S.~M., Kaufmann, T., 
       \& Wadsley, J.\ 2009, \mnras, 395, 1455   
\bibitem[Stringer \& Benson(2007)]{Stringer07} Stringer, M.~J., \& Benson, A.~J.\ 2007, \mnras, 382, 641
\bibitem[Stringer et al.(2011)]{Stringer+11} Stringer, M., Cole, 
S., Frenk, C.~S., \& Stark, D.~P.\ 2011, \mnras, 481  %in press, arXiv:1011.2745 
\bibitem[Tonini et al.(2009)]{2009MNRAS.396L..36T} Tonini, C., Maraston, C., Devriendt, J., Thomas, D., \& Silk, J.\ 2009, \mnras, 396, L36
\bibitem[Valenzuela et al. (2007)]{Valenzuela07} Valenzuela, O., Rhee, G., Klypin, A., Governato, F., Stinson, G., 
Quinn, T., \& Wadsley, J., 2007, \apj, 657, 773
\bibitem[van den Bosch(2000)]{vandenBosch00} van den Bosch, F.~C.\ 2000, \apj, 530, 177
\bibitem[Vergani et al.(2008)]{Vergani08} Vergani, D., et al.\ 2008, \aap, 487, 89	
\bibitem[Wang \& Jing(2010)]{WangJing10} Wang, L., \& Jing, Y.~P.\ 2010, \mnras, 402, 1796 
\bibitem[Weinmann et al.(2011)]{Weinmann11} Weinmann, S.~M., van den Bosch, F.~C., \& Pasquali, A.\ 2011, 
arXiv:1101.3244 
\bibitem[White \& Rees(1978)]{WR78} White, S.~D.~M., \& Rees, M.~J.\ 1978, \mnras, 183, 341 
\bibitem[White \& Frenk(1991)]{WF91} White, S.~D.~M., \& Frenk, C.~S.\ 1991, \apj, 379, 52 
\bibitem[Wuyts et al.(2008)]{Wuyts08} Wuyts, S., Labb{\'e}, I., 
Schreiber, N.~M.~F., Franx, M., Rudnick, G., Brammer, G.~B., \& van Dokkum, P.~G.\ 2008, \apj, 682, 985 
\bibitem[Yang et al.(2009)]{Yang09} Yang, X., Mo, H.~J., \& van den Bosch, F.~C.\ 2009, \apj, 695, 900
\bibitem[Zavala et al.(2008)]{Zavala08} Zavala, J., Okamoto, T.,   \& Frenk, C.~S.\ 2008, \mnras, 387, 364 
\bibitem[Zheng et al. (2007)]{Zheng07} 
         Zheng X.~Z., Bell E.~F., Papovich C., Wolf C., Meisenheimer K., Rix H.-W., 
         Rieke G.~H., Somerville R., 2007, ApJ, 661, L41
         
\end{thebibliography}
\end{document}